\documentclass{article}
\usepackage[utf8]{inputenc}
\usepackage{authblk}
\usepackage{amsmath}
\usepackage{amssymb}
\usepackage{enumitem}
\usepackage{caption}
\usepackage{mathtools}
\usepackage{subcaption}
\usepackage[table]{xcolor}
\usepackage[boxruled,linesnumbered]{algorithm2e}
\usepackage{optidef}
\usepackage{hyperref}
\usepackage{float}
\usepackage{natbib}
\usepackage{bm,bbm}
\usepackage{tikz}
\usetikzlibrary{arrows,shapes,positioning,shadows,trees}
\restylefloat{table}
\title{Precision versus Shrinkage: A Comparative Analysis of Covariance Estimation Methods for Portfolio Allocation}
\author[1]{Sumanjay Dutta}
\author[1]{Shashi Jain}
\affil[1]{Department of Management Studies, Indian Institute of Science} 
\date{}
\begin{document}
\maketitle
\begin{abstract}
     In this paper, we perform a comprehensive study of different covariance and precision matrix estimation methods in the context of minimum variance portfolio allocation. The set of models studied by us can be broadly categorized as: Gaussian Graphical Model (GGM) based methods, Shrinkage Methods, Thresholding and Random Matrix Theory (RMT) based methods. Among these, GGM methods estimate the precision matrix directly while the other approaches estimate the covariance matrix. We perform a synthetic experiment to study the network learning and sample complexity performance of GGM methods. Thereafter, we compare all the covariance and precision matrix estimation methods in terms of their predictive ability for daily, weekly and monthly horizons. We consider portfolio risk as an indicator of estimation error and employ it as a loss function for comparison of the methods under consideration. We find that GGM methods outperform shrinkage and other approaches. Our observations for the performance of GGM methods are consistent with the synthetic experiment. We also propose a new criterion for the hyperparameter tuning of GGM methods. Our tuning approach outperforms the existing methodology in the synthetic setup.  We further perform an empirical experiment where we study the properties of the estimated precision matrix. The properties of the estimated precision matrices calculated using our tuning approach are in agreement with the algorithm performances observed in the synthetic experiment and the empirical experiment for predictive ability performance comparison. Apart from this, we perform another synthetic experiment which demonstrates the direct relation between estimation error of the precision matrix and portfolio risk.
    \end{abstract}
 
\section{Introduction} \label{intro}
  The estimation of the covariance matrix of stock returns constitutes a crucial aspect of the portfolio allocation process. Conventionally, this involves compiling a historical data set of stock returns and computing the sample covariance matrix. However, it has been observed that this technique poses challenges in terms of accurate estimation. Specifically, when the number of stocks being evaluated exceeds the available historical return observations, the sample covariance matrix is prone to substantial estimation errors, which can have a disproportionate impact on the entries with high correlations. Consequently, pairs with extreme entries may not accurately reflect the underlying relationships among the stocks, but rather reflect the high levels of errors associated with their estimation. Invariably, the mean-variance optimization algorithm tends to prioritize the coefficients with the highest levels of estimation error and allocate its largest investments accordingly. The reason for this is that coefficients with the highest estimation error often have more extreme values, and as a result, they are given higher weights by the mean-variance optimization algorithm, leading to larger investments being allocated to them, even though they may not be reliable.
 The critiques related to estimation noise can be categorized into two key components: 
\begin{enumerate}
\item  the estimate of the sample covariance matrix, and 
\item the inverse of the sample covariance matrix, also called the Precision Matrix. 
\end{enumerate} 
The first problem is tackled by different methods  broadly centred around three approaches: a) \emph{Shrinkage},  b) \emph{Thresholding} and c) \emph{Random Matrix Theory} (RMT) Methods.
 Shrinkage tends to pull the most extreme coefficients towards more central values, thereby systematically reducing estimation error where it matters most. It requires three elements: An estimator with no structure, an estimator with a lot of structure, and a shrinkage constant. The estimator without structure is generally quite obvious, given the context. For us it is the sample covariance
matrix. Statistically, the challenge is to know the structured estimator and the optimal shrinkage constant.

 The utilization of shrinkage methods for portfolio selection was initially explored by \cite{frost1986empirical} and \cite{jorion1986bayes}. However, their specific techniques for shrinkage were unable to handle cases where the number of assets surpassed the quantity of historical return observations, a common occurrence in practical applications. Much of the foundations for the Shrinkage approach for covariance estimation has been laid out by Ledoit and Wolf (\cite{ledoit2004honey}, \cite{ledoit2004well}), \cite{schafer2005shrinkage} and \cite{chen2010shrinkage} have suggested other shrinkage techniques which can be used to get cleaner covariance matrix estimates.
  \cite{jagannathan2003risk} proved that mean-variance optimizers are already implicitly applying some form of shrinkage to the sample covariance matrix when short sales are ruled out, and this improves the stability of the weights.

Thresholding involves setting small values in the sample covariance matrix to zero, effectively \emph{shrinking} the matrix towards a simpler structure with fewer parameters. The threshold value is usually chosen based on a statistical criterion, such as the universal threshold or the adaptive threshold.The idea behind thresholding is that small values in the sample covariance matrix may be due to noise or sampling error, and are not reflective of the true underlying covariance structure. By setting these values to zero, the estimate becomes more stable and accurate, particularly in high-dimensional settings.   \cite{donoho1995wavelet}, \cite{bickel2008covariance} and \cite{cai2011adaptive} have proposed thresholding techniques that estimate the covariance matrix. 

Another approach, which comes from Random Matrix Theory (RMT) attempts to clean the measurement noise in the covariance and correlation matrices. RMT is used in the study of correlation matrices through identification of the eigenvalue distribution of large correlation matrices. The eigenvalues of a correlation matrix are related to the principal components of the data, which capture the major sources of variation in the dataset. RMT provides a theoretical framework for understanding the behavior of the eigenvalues of large random correlation matrices. By analysing the empirical eigenvalue distribution of a correlation matrix, RMT methods can identify the presence of significant correlations between variables. \cite{bun2017cleaning} discuss several RMT methods and analytical techniques (see \cite{bun2018optimal}, \cite{ledoit2011eigenvectors}). They also empirically establish the efficacy of the Rotationally Invariant Estimator (RIE) method (see \cite{bun2016rotational}). These RMT methods provide cleaned correlation matrices, which can then used to derive both the covariance and precision matrices.
 
   To circumvent the second critique, it is possible to estimate the precision matrix directly, rather than inverting the covariance matrix. Thus, a need for alternative techniques of precision matrix estimation arises which may reduce the estimation error and provide better out-of-sample performance. Gaussian Graphical Model (GGM) based techniques have been proposed to estimate the precision matrix directly. GGM methods  were shown to provide consistent estimates of the precision matrix.  \cite{friedman2008sparse}, (\cite{cai2011constrained} and \cite{meinshausen2006high}) propose GGM methods which provide Sparse and consistent estimates of the Precision Matrix. \cite{millington2017robust} conducted an empirical study that applies Graphical Lasso for the estimation of covariance for the portfolio allocation. Their empirical findings suggest that portfolios that use Graphical Lasso for covariance estimation yield higher returns and have lower risk compared to the empirical covariance matrix. \cite{torri2019sparse} propose a data-driven portfolio framework based on two regularization methods, the Graphical Lasso (\textit{Glasso}) and \textit{tlasso}, that provide sparse estimates of the precision matrix by penalizing its $L_{1}$-norm. \cite{callot2021nodewise} use the nodewise-regression method of \cite{meinshausen2006high} to establish consistency of the estimated variance, weights and risk of high-dimensional financial portfolio. Their empirical application demonstrates that the precision matrix estimator based on the nodewise-regression outperforms the principal orthogonal complement thresholding estimator (POET) of \cite{fan2013large} and linear shrinkage estimator of \cite{ledoit2004well}. \par
 The estimation methods compared in this study have been broadly classified in figure\ref{f1}:
\begin{figure}[!h]
\centering
\tikzset{
  basic/.style  = {draw, text width=2cm, drop shadow, font=\sffamily, rectangle},
  root/.style   = {basic, rounded corners=2pt, thin, align=center,
                   fill=green!30},
  level 2/.style = {basic, rounded corners=6pt, thin,align=center, fill=green!60,
                   text width=8em},
  level 3/.style = {basic, thin, align=left, fill=pink!60, text width=6.5em}
}

\begin{tikzpicture}[
  level 1/.style={sibling distance=60mm},
  edge from parent/.style={->,draw},
  >=latex]

\node[root] {Estimation Methods for Low-Sample Regimes}
  child {node[level 2] (c1) {Estimate the Covariance Matrix}}
  child {node[level 2] (c2) {Estimate the Inverse Covariance Matrix}};

\begin{scope}[every node/.style={level 3}]
\node [below of = c1, xshift=15pt] (c11) {Shrinkage};
\node [below of = c11] (c12) {Thresholding};
\node [below of = c12] (c13) {Random Matrix Theory};

\node [below of = c2, xshift=15pt] (c21) {GGM Methods};

\end{scope}

\foreach \value in {1,...,3}
  \draw[->] (c1.195) |- (c1\value.west);

\foreach \value in {1}
  \draw[->] (c2.195) |- (c2\value.west);
\end{tikzpicture}
 \caption{Classification of Estimation Methods}
 \label{f1}
 \end{figure}
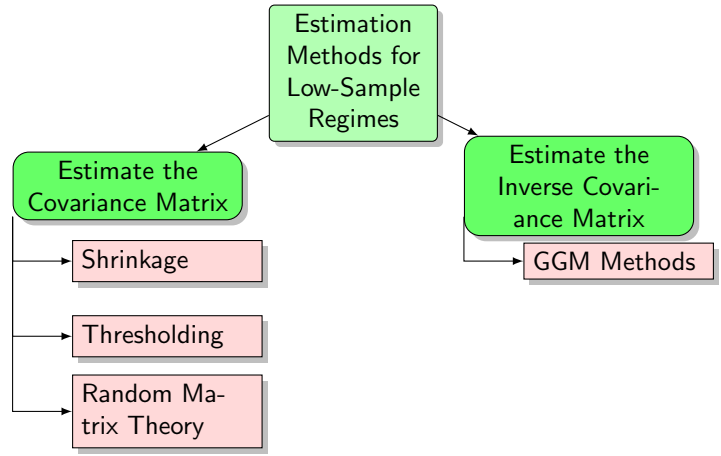
 
  We study a wide class of methods in this paper in terms of their predictive ability for the minimum variance portfolio allocation problem. This raises the question of providing a statistical procedure that delivers the \emph{best} performing methods with respect to a given criterion. Sometimes it is not trivial to assess which model clearly outperforms each other. This problem is relevant from an empirical point of view especially when the set of competing alternatives is large, as in our context. Recently, a lot of effort has been devoted to develop new testing procedures being able to deliver the \emph{best fitting} models, see e.g. the Reality Check (RC) of \cite{white2000reality}, the Stepwise
Multiple Testing procedure of \cite{romano2005stepwise} and the Superior Predictive Ability (SPA) test of \cite{hansen2005test}. \cite{jain2019can} use the SPA test to  assess whether the machine learning based portfolio allocation models outperform the traditional risk based portfolios. The portfolio variance is directly proportional to the estimation error and can be used as a loss function to compare models. We compare these losses to rank the performance of the competing methods using the Model Confidence Set (MCS) Procedure proposed by \cite{hansen2011model}, which consists on a sequence of tests which permits to construct a set of \emph{superior} models, where the null hypothesis of Equal Predictive Ability (EPA) is not rejected at a certain confidence level. 

 We address the following problems in this paper: 
\begin{itemize}
\item We first study the performance of the GGM methods in terms of structure recovery and sample complexity in a synthetic setup. We control the behaviour of the data by imposing a Multivariate Gaussian structure. The question  we therefore address is: Do GGM methods exhibit a relationship between sample complexity for network learning and the predictive ability for portfolio allocation ? 
\item Second, we study the properties of the estimated precision matrices recovered using GGM methods. This give us insights about the sparsity, error  sensitivity and network properties of the estimates. We find that stock returns exhibit an important network property, which makes them suitable for direct precision matrix estimation through GGM methods.
\item Third, we perform an empirical study which compares the four broad classes of estimation methods mentioned earlier in terms of the minimum variance portfolio allocation problem. Predictive ability is directly associated with the  measurement errors of the estimated covariance and precision matrices. Therefore, we analyse if it possible to rank these methods in terms of predictive ability. This also provides us a view of how the estimation errors associated with these methods impact the portfolio allocation process. We employ the MCS procedure to rank the methods. For this, we use the daily and intra-day returns data of the constituent stocks of the Nifty500 index, which is a representative of the top 500 stocks (by market capitalization) listed at the National Stock Exchange (NSE), India.
\item Apart from this, we also compare the best performing precision and covariance matrix estimation methods in another synthetic setup where we demonstrate the connection between precision matrix estimation error and predictive ability.
\item  We also study the impact of the choice of criterion for parameter tuning  on the performance of GGM methods for portfolio allocation. We propose  Portfolio variance as the objective function to be minimized as opposed to the approach suggested by \cite{kelner2020learning} for hyperparameter selection. This criterion is used in both the synthetic and empirical setups and has important implications for the observations recorded.
\item The literature on GGM methods uses the finite grid parameter choice approach to estimate the precision matrix. This makes the performance of the methods dependent on a rather arbitrary choice of grids. In order to address this problem, we propose a novel procedure of parameter choice based on the Nelder-Mead Optimization method, which allows us to extended our choice of parameters from a finite sized discrete grid to a global continuous grid thus allowing for a wider choice of parameters for estimation.  
\end{itemize}

Observations from our empirical experiments suggested that GGM methods clearly outperform the covariance matrix estimation methods. We also rank the covariance matrix estimation methods separately and find that Non-linear shrinkage approach performs better than linear shrinkage methods. Overall, shrinkage methods also outperform the thresholding approach. The RMT based method is comparable with some shrinkage methods. Apart from this, our tuning approach provides better algorithm performance in terms of sample complexity for network learning for GGM methods. This is also substantiated by the empirical experiment, where our tuning approach yields the best performing methods. Our second synthetic experiment also suggests that precision matrix estimation error is directly linked with predictive ability. That is, the higher the estimation error, the lower the predictive ability.    

The paper is organized into the following sections: Section \ref{Models} gives a brief description of the methods under discussion. It also discusses the hyperparameter tuning procedures used for these methods. Section \ref{SPA and MCS} discusses the statistical testing procedures employed for comparing these methods in terms of predictive ability.  Section \ref{Synthetic Experiment} discusses the design and results for the synthetic experiment. Section \ref{Data and Methodology} presents the design for our empirical analyses. Section \ref{Emp exp} presents and discusses the results observed from our empirical analyses. Section \ref{Conclusion} discusses the conclusions from our experiments.  The Appendices \ref{Ap1} and \ref{Ap2} discuss the issue of covariance matrix estimation error using spectral decomposition and the Nelder-Mead style approach for grid extension respectively. Appendix \ref{Ap3} lists all models considered for analysis in this paper. 
\section{Precision and Covariance Matrix Estimation Methods} \label{Models} 
Since the seminal work of \cite{markowitz77portfolio}, the idea of the minimization of  risk through portfolio diversification has become central to  portfolio theory. The idea behind this portfolio allocation model is simple: One needs to estimate the covariance matrix from the returns of the individual components of the portfolio and thus derive an analytical solution via a simple quadratic optimization problem. This is possibly among the key factors behind the widespread usage of the Markowitz framework, which is stated below:

\begin{mini}|l|
   {w}{w^{\top} \Sigma w}{}{}
   \addConstraint{1^{\top} w }{= 1,}{}
\end{mini}
where $\Sigma$ represents the $n \times n$ true covariance matrix, $w$ is the $n \times 1$ vector of asset
weights and $1$ denotes a $n \times 1$ unit vector. Asset returns are used to compute the covariance matrix.  The analytical solution for the optimization problem is as follows:
\begin{equation}
    w_{MV} = \frac{\Sigma^{-1}1}{1^{\top}\Sigma^{-1}1}
\end{equation}
where $w_{MV}$ is the vector of weights of the global minimum variance portfolio. In order to obtain the weights, an estimate $\hat{\Sigma}$ has to be computed (since $\Sigma$ is unknown):
\begin{equation} 
     \hat{w}_{MV} = \frac{\hat{\Sigma}^{-1}1}{1^{\top}\hat{\Sigma}^{-1}1}
\end{equation}
As discussed before, computation of the sample covariance and its inverse is prone to estimation errors. \cite{meucci2005risk} shows that this approach leads to portfolio weights with poor out-of-sample properties. In this paper, we use shrinkage, thresholding and Random Matrix methods to supply $\hat{\Sigma}$. GGM methods, on the other hand, are used to estimate $\hat{\Sigma}^{-1}$. We compare these two strands of estimation in terms of their error reduction efficiency while solving the global minimum variance portfolio allocation problem. 
\subsection{Gaussian Graphical Models} \label{GGMs}
The literature on precision matrix estimation for the low sample regime has made use of GGM methods for avoiding an inversion of the covariance matrix. This helps in circumventing the estimation losses associated with such an inversion. A GGM in $n$ dimensions is a probability distribution with density
\[p(X = x) = \frac{1}{\sqrt{(2\pi)^{n} \lvert \Sigma \rvert}} \textnormal{exp}\left(-(x-\mu)^{T}\Sigma^{-1}(x-\mu)/2 \right)\]
where $\mu$ is the mean and $\Sigma$ is the covariance matrix. The
important point is that their conditional independence structure is encoded by $\Theta = \Sigma^{-1}$. We can associate a graph to $\Theta$ which connects two nodes $i$, $j$ when $\Theta_{ij} \neq 0$. Now each node $i$ only interacts directly with its neighbors in the sense that $X_{i}$ is conditionally
independent of every other node in the graphical model given its neighbors $(X_{j} )_{i \sim j}$. This means that there is a directly connection between learning the network structure and the estimation error of the precision matrix. We study this link more closely in our synthetic and empirical experiments. In the context of portfolio optimization, $X$ denotes the returns matrix of the correlated assets. 

GGM methods for precision matrix estimation can be categorized into two approaches. The first approach for using GGM methods to estimate the precision matrix focuses on constraining the $L_{1}$ norm, where the goal is to learn the sparsity pattern of $\Theta$ assuming some sort of lower bound on the strength of non-zero interactions. This is a natural objective because once the sparsity pattern is known, estimating the entries of $\Theta$ is straightforward (e.g. one can use ordinary least
squares). Glasso, Clime and MB (discussed in \ref{Glasso}, \ref{Clime} and \ref{MB}) belong to this category. 

 While the literature concerning the first approach shows that sparse GGMs can be estimated when the number of samples is logarithmic
in the dimension, there is an important caveat in their guarantees. They need to assume that $\Theta$ is in some sense well-conditioned. In response to this, the second approach was proposed by \cite{kelner2020learning}. This was motivated by \cite{misra2020information} who studied the question of how many samples are needed information-theoretically to learn sparse GGMs in the ill-conditioned case. \cite{kelner2020learning} show that  relying on minimizing the conditional variance per node for estimating the precision matrix leads to better sample complexity (that is, these algorithms learn the partial correlation structure of the network with lesser number of samples) and higher sparsity. \cite{kelner2020learning} propose the Greedy Prune and HybridMB, which use this approach. Greedy Prune and HybridMB study two important classes of GGMs respectively, viz. \emph{Attractive} and \emph{Walk-Summable models}. A GGM is called attractive if $\Theta_{ij} \leq 0$ for all $i \neq j$. \\* On the other hand,  walk-summable matrices are defined as follows: $\Theta$ is walk-summable if making all off-diagonal entries of $\Theta$ negative preserves the fact that $\Theta$ is positive definite. All attractive GGMs are walk-summable, as are other important classes of GGMs like pairwise normalizable and non-frustrated models (\cite{malioutov2006walk}). The GGM methods are summarily classified below in figure \ref{f2} :
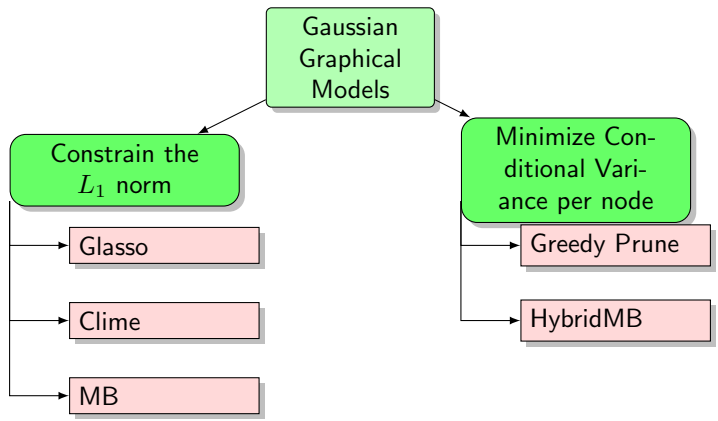
\begin{figure}[!h]
\centering
\tikzset{
  basic/.style  = {draw, text width=2cm, drop shadow, font=\sffamily, rectangle},
  root/.style   = {basic, rounded corners=2pt, thin, align=center,
                   fill=green!30},
  level 2/.style = {basic, rounded corners=6pt, thin,align=center, fill=green!60,
                   text width=8em},
  level 3/.style = {basic, thin, align=left, fill=pink!60, text width=6.5em}
}

\begin{tikzpicture}[
  level 1/.style={sibling distance=60mm},
  edge from parent/.style={->,draw},
  >=latex]

\node[root] {Gaussian Graphical Models}
  child {node[level 2] (c1) {Constrain the $L_{1}$ norm}}
  child {node[level 2] (c2) {Minimize Conditional Variance per node}};

\begin{scope}[every node/.style={level 3}]
\node [below of = c1, xshift=15pt] (c11) {Glasso};
\node [below of = c11] (c12) {Clime};
\node [below of = c12] (c13) {MB};

\node [below of = c2, xshift=15pt] (c21) {Greedy Prune};
\node [below of = c21] (c22) {HybridMB};

\end{scope}

\foreach \value in {1,...,3}
  \draw[->] (c1.195) |- (c1\value.west);

\foreach \value in {1,...,2}
  \draw[->] (c2.195) |- (c2\value.west);
\end{tikzpicture}
 \caption{Classification of Gaussian Graphical Models}
 \label{f2}
 \end{figure}
\newline
We provide a brief description of the GGM methods employed in our analyses:
\subsubsection{Graphical Lasso (Glasso)}\label{Glasso}
 Suppose we have $N$ multivariate normal observations of dimension $p$, with
mean $\mu$ and covariance $\Sigma$. Following \cite{friedman2008sparse}, let $\Theta = \Sigma^{-1}$ and let $\hat{\Sigma}$ be the sample covariance matrix, the problem is to maximize the log-likelihood
\[\textnormal{log det} \Theta - \textnormal{tr}(\hat{\Sigma}\Theta) - \rho||\Theta||_{1}\]
over non-negative definite matrices $\Theta$ Here tr denotes the trace and $||\Theta||_{1}$
is the $L_{1}$ norm- the sum of the absolute values of the elements of $\Sigma^{-1}$. This expression is the Gaussian log-likelihood of the data, partially maximized with respect to the mean parameter $\mu$.
Since we are interested in settings where the number of samples is much smaller than the dimension, $\hat{\Sigma}$ is a  somewhat crude approximation to the true covariance. However, it is a good estimate when restricted to
sparse directions.   \newline
\cite{awoye2016markowitz} used Glasso (\cite{friedman2008sparse}) to estimate a sparse covariance matrix for the Markowitz mean-variance portfolio problem to improve covariance estimation in terms of lower realized portfolio risk. 
\subsubsection{Meinhausen-B\"{u}hlman}\label{MB}
\cite{meinshausen2006high} show that neighborhood selection with the Lasso is a computationally attractive alternative to standard covariance selection for sparse high-dimensional graphs. Their method estimates an unknown precision matrix using the idea of projections to approximately invert the sample covariance matrix.  For each $j = 1, . . . , p$ they define the vector $\gamma = \{\gamma_{j,k}, j \neq k\}$ as follows:
\[
\gamma_{j} := \min_{\gamma \in \mathbb{R}^{p-1}}\mathbb{E}||X_{j} - X_{-j\gamma}||^{2}_{2}/n \]
and denote $\eta_{j} := X_{j} - X_{-j\gamma}$ and the noise level by $\tau_{j}^{2} = \mathbb{E}\eta^{T}_{j}\eta_{j}/n$. For the column vector defined as: $\Gamma_{j} := (-\gamma_{j,1},\ldots, -\gamma_{j,j-1}, 1, -\gamma_{j,j+1}, -\gamma_{j,p})^{T}$,  they show that
\[\Theta_{0}= (\Theta_{1}^{0},\ldots,\Theta_{p}^{p}) = (\Gamma_{1}/\tau_{1}^{2},\ldots,\Gamma_{p}/\tau_{1}^{p}) \]
where $\Theta^{0}_{j}$ is the j-th column of $\Theta_{0}$. Hence the precision matrix $\Theta_{0}$ may be recovered
from the partial correlations $\gamma_{j,k}$ and from the noise level $\tau^{2}_{j}$.
\cite{callot2021nodewise} use the nodewise-regression method of \cite{meinshausen2006high} to establish consistency of the estimated variance, weights and risk of high-dimensional financial portfolio. Their empirical application demonstrates that the precision matrix estimator based on the nodewise-regression outperforms the principal orthogonal complement thresholding estimator
(POET) (\cite{fan2013large}) and linear shrinkage (Ledoit and Wolf (2004)\cite{ledoit2004well}). 
\subsubsection{Clime}\label{Clime}
Another popular approach is the Clime estimator \cite{cai2011constrained} which solves the following linear program
\[\textnormal{min} ||\Theta||_{1} \\\ \textrm{s.t.} \\\ |\Sigma_{n}\Theta - I|_{\infty} \leq \lambda_{n}\] The analysis of Clime assumes a bound $M$ on the maximum $L_{1}$-norm of any row of the inverse covariance (given that the $X_{i}$’s are standardized to unit variance). This is also a type of condition number assumption, although of a different nature than RE \cite{kelner2020learning} and \cite{friedman2008sparse}.  \cite{cai2020high} use Clime to develop a consistent estimator of the minimum variance for high-dimensional global minimum-variance portfolio. 
\subsubsection{Greedy Prune}\label{Greedy Prune} 
\cite{kelner2020learning} introduce the the following algorithm and show that it succeeds in learning the graph structure of attractive GGMs. This algorithm, called Greedy Prune, does the following to learn the neighborhood of node $i$:
\begin{enumerate}
    \item Set $S = \emptyset$ (S is the set of nodes not containing all neighbors of $i$)  and let $\nu > 0$ be a thresholding parameter.
    \item (Greedy/ step) Repeat the following $T$ times: set $j$ to be the the minimizer of $Var( X_{i}|X_{S}, X_{j} )$ and add $j$ to $S$.
    \item (Pruning step) For each $j \in S$: if $\hat{Var}(X_{i}
|X_{S}) > (1 - \nu)\hat{Var}(X_{i}
|X_{S \ \{j\}})$, remove $j$ from $S$.
\item Return $S$ as the neighborhood of node $i$.
\end{enumerate}
 where $\hat{Var}$ indicates the variance is estimated from sample, using Ordinary Least Squares. 
\subsubsection{Hybrid MB}\label{HybridMB}

The HybridMB algorithm (see \cite{kelner2020learning} aims at learning walk-summable GGMs. This algorithm does the following to learn the neighborhood of node $i$:
\begin{enumerate}
    \item (Greedy step) Set $j$ to be the minimizer of $\hat{Var}\left(X_{i}|X_{j}\right)$.
    \item (Lasso with implicit weak preconditioning) Solve for $w$, $a$ in
    \[
\begin{aligned}
\min_{w, a:||w|| \leq \lambda} \quad & \hat{E} \left[\left(X_{i} - \sum_{k \notin \{i,j\}}w_{k}\frac{X_{k}}{\sqrt{\hat{Var}(X_{k}|X_{j}}} - aX_{j}\right)^{2} \right]\\
\end{aligned}
\]
   \item (Pruning step) We perform a pruning step similar to Greedy Prune to zero out some of the entries of $w$, and to test if $j$ is an actual neighbor.
\item  Return $j$ (if it passed the test) and the remaining support of $w$ as the neighborhood of $i$.
\end{enumerate}
\subsection{Shrinkage Methods}
Suppose we have a sample of $n$ $\mathrm{iid}$ random vectors in $p$ dimensions, with an expected value of zero. The $p \times p$ covariance matrix can be estimated by the usual sample covariance matrix (SCM). From eigenvalue decomposition, we know that $\Sigma = U \Lambda U'$, with $U$ being orthogonal. $\Lambda$ is a diagonal matrix containing the eigenvalues of the covariance matrix. In particular, we see that the covariance matrix is invertible if and only if all eigenvalues are different from zero, or in other words, if the rank of the matrix is $p$.

However in the above covariance matrix, we have $p*(p-1)/2$ elements to estimate. $p*(p-1)$ is the number of ways to choose two elements out of $p$, and then we divide by 2 because the matrix is symmetric. This becomes a problem if $n$ is not much larger than $p$! This is observed when $p > n$ occurs. In this case, it can be shown that the SCM has a rank of at most $n < p$. In particular, even if the true covariance matrix is invertible (having all eigenvalues larger than zero), the SCM will have $p-n$ zero eigenvalues and thus never be invertible. This might be extreme as an extreme case, but even for a case like $p=n/2$, the estimation error of the SCM can be substantial. 
\newline
\newline
\newline
\textbf{Linear Shrinkage Methods} \newline
An arguably simple idea that has led to a myriad of research and applications is to just take a linear combination of SCM and the identity matrix:
\[\hat{\Sigma} = (1-a)\hat{S} + aI\]
The shrinkage constant $a$ can be chosen according to  theoretical considerations. The different approaches to Linear Shrinkage vary in their methodology for choosing $a$. The Ledoit Wolf Linear Shrinkage estimator, Rao-Blackwell Ledoit Wolf estimator, Bodnar's Shrinkage estimator and Oracle Approaching approaches (see \ref{LWL}, \ref{RBLW}, \ref{BDL} and \ref{OA}) belong to this class. \newline
\newline
\textbf{Nonlinear Shrinkage Methods:} \newline
Nonlinear shrinkage derives an asymptotic estimator in a much larger class (that need not be just a linear function of the SCM). In particular, it solves the problem
\[\arg \min_{\Lambda} l(\hat{\Sigma},\hat{U}\Lambda\hat{U}') \]
where $l$ is some loss function.We are looking for the choice of (shrunken) eigenvalues that brings the resulting matrix as close as possible to the true covariance matrix if we are only allowed to use the eigenvectors of the SCM. As we have seen before linear shrinkage is a special case of this because we also only change the eigenvalues there. However, now the way we determine Lambda is not constrained to be linear, offering much more flexibility.We include the Analytical Non-Linear Shrinkage Estimator (see \ref{LWNL}) proposed by \cite{ledoit2020analytical} in our study. We present a summarized classification of the Shrinkage Approach in figure \ref{f3}:
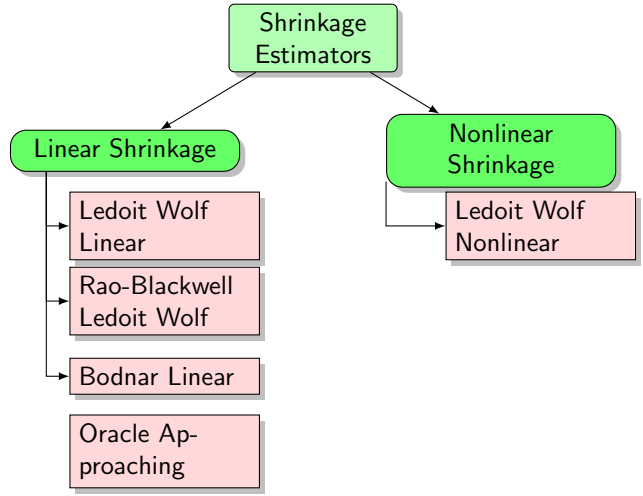
\begin{figure}[!h]
\centering
\tikzset{
  basic/.style  = {draw, text width=2cm, drop shadow, font=\sffamily, rectangle},
  root/.style   = {basic, rounded corners=2pt, thin, align=center,
                   fill=green!30},
  level 2/.style = {basic, rounded corners=6pt, thin,align=center, fill=green!60,
                   text width=8em},
  level 3/.style = {basic, thin, align=left, fill=pink!60, text width=6.5em}
}

\begin{tikzpicture}[
  level 1/.style={sibling distance=50mm},
  edge from parent/.style={->,draw},
  >=latex]

\node[root] {Shrinkage Estimators}
  child {node[level 2] (c1) {Linear Shrinkage}}
  child {node[level 2] (c2) {Nonlinear Shrinkage}};

\begin{scope}[every node/.style={level 3}]
\node [below of = c1, xshift=15pt] (c11) {Ledoit Wolf Linear};
\node [below of = c11] (c12) {Rao-Blackwell Ledoit Wolf};
\node [below of = c12] (c13) {Bodnar Linear};
\node [below of = c13] (c14) {Oracle Approaching};

\node [below of = c2, xshift=15pt] (c21) {Ledoit Wolf Nonlinear};

\end{scope}

\foreach \value in {1,...,3}
  \draw[->] (c1.195) |- (c1\value.west);

\foreach \value in {1}
  \draw[->] (c2.195) |- (c2\value.west);
\end{tikzpicture}
 \caption{Classification of Shrinkage Estimators}
 \label{f3}
 \end{figure}

\subsubsection{Ledoit Wolf Linear Shrinkage Estimator}\label{LWL}
Ledoit and Wolf (2004) \cite{ledoit2004well} proposed a linear shrinkage strategy (LWL) to estimate covariance matrix with an application to portfolio optimization. An optimal covariance is written as a convex combination as follows,
\[\hat{\Sigma} = \delta\hat{F} + (1 - \delta)S\]
where $\delta \in (0, 1)$ is a control parameter/weight, $\hat{S}$ an empirical covariance matrix, and $\hat{F}$ a target matrix. This paper introduces an
estimator that is both well-conditioned and more accurate than the sample covariance matrix asymptotically. This estimator is distribution-free and has a simple explicit formula that is easy to compute and interpret. It is the asymptotically optimal convex linear combination of the sample covariance matrix with the identity matrix. 

\subsubsection{Rao-Blackwell Ledoit Wolf Estimator}\label{RBLW}
\cite{chen2010shrinkage} (RBLW) address covariance estimation in the sense of minimum mean-squared error (MMSE) for Gaussian samples. Specifically, we consider shrinkage methods which are suitable for high dimensional problems with a small number of samples (large p small n). They modify the Ledoit-Wolf (LW) method by conditioning on a sufficient statistic. By the Rao-Blackwell theorem, this yields the new estimator. This approach estimates covariance matrix by minimizing mean squared error with the following formula,
\[\hat{\Sigma} = \rho \hat{F} + (1 - \rho)\hat{S}\]
where $\rho \in (0, 1)$ a control parameter/weight, $\hat{S}$ an empirical covariance matrix, and $\hat{F}$ a target
matrix. It is proposed to use a structured estimate $\hat{F} = Tr(\hat{S}/p ).I_{p\times p}$ where $I_{p\times p}$ is an identity
matrix of dimension $p$. 
\subsubsection{Oracle Approaching Shrinkage Estimator}\label{OA}
\cite{chen2010shrinkage} propose an iterative approach (OA) which approximates the clairvoyant shrinkage estimator. Convergence of this iterative method is established and a closed form expression for the limit is determined.
The authors estimate the covariance matrix by iteratively approximating the shrinkage with
\[\hat{\Sigma} = \rho \hat{F} + (1 - \rho)\hat{S} \]
where $\rho \in (0, 1)$  is a control parameter/weight, $\hat{S}$ is an empirical covariance matrix, and $\hat{F}$ a target
matrix. It is proposed to use a structured estimate $\hat{F} = Tr(\hat{S}/p ).I_{p\times p}$ where $I_{p\times p}$ is an identity
matrix of dimension $p$.
\subsubsection{Bodnar's Linear Estimator}\label{BDL}
In this paper \cite{bodnar2014strong}  extend the work of \cite{ledoit2004well} by constructing a more general linear shrinkage estimator (BDL) for
a large dimensional covariance matrix. The target matrix here is considered to be an arbitrary symmetric positive definite
matrix with uniformly bounded trace norm. The shrinkage estimator is given by 
\[\hat{\Sigma_{OLSE}} = \hat{\alpha}S_{n} + \hat{\beta}\Sigma_{0} \]
where $\hat{\alpha}$ and $\hat{\beta}$ are the estimated shrinkage intensities. The recent results from the random matrix theory allow them to find the asymptotic deterministic equivalents of the optimal shrinkage intensities and estimate them consistently. The developed distribution-free estimators obey almost surely the smallest Frobenius loss over all linear shrinkage estimators for the covariance matrix. 
\subsubsection{Ledoit Wolf Analytical Non-linear Estimator}\label{LWNL}
 \cite{ledoit2020analytical} provide an analytical formula for optimal nonlinear
shrinkage (LWNL) of large-dimensional covariance matrices. It is achieved by identifying and mathematically exploiting a deep connection between nonlinear shrinkage and non-parametric estimation of the Hilbert transform of the sample spectral density.The nonlinear shrinkage estimator of the covariance matrix is the one minimizes the minimum variance:
loss function
\[L^{MV}_{n} = \frac{Tr(\hat{\Sigma}^{-1}_{n}\Sigma_{n}\hat{\Sigma}^{-1}_{n})/p}{[Tr(\hat{\Sigma}^{-1})/p]^{2}} - \frac{1}{Tr(\hat{\Sigma}^{-1})/p}\]
where $\textrm{Tr(·)}$ denotes the trace of a square matrix, $\Sigma_{n}$ represents the estimated covariance matrix and $L^{MV}_{n}$ represents the true variance of the linear combination of the original variables that has the minimum estimated variance, under a generic linear constraint, after suitable normalization. They also show that their method is faster than previous nonlinear shrinkage methods, which were numerical.
\subsection{Thresholding Estimators}
\cite{bickel2008covariance} proposed thresholding of the sample covariance matrix $\Sigma$ and obtained rates of convergence for the thresholding estimators. This approach is similar to the Shrinkage Approach (which relies on regularization of the Covariance Matrix) except for the fact that it specifically deals with \emph{bandable} matrices, where the entries of the matrix decay as they move away from the diagonal. Similar approaches have been used by \cite{antoniadis2001regularization} and \cite{cai2011adaptive}. We discuss some of these thresholding methods below. 
\subsubsection{Hard Thresholding Estimator}
 \cite{bickel2008covariance} proposed a sparse covariance estimation technique (HT) to apply thresholding on off-diagonal elements of the sample covariance matrix. The entry of sample covariance matrix $S_{i,j} = 0$ if $|S_{i,j}| \leq \tau$ where $\tau$ is a thresholding value). If the thresholding value is rather a vector of
regularization parameters, it applies cross-validation scheme to select an optimal value. They also discuss an intuitive resampling scheme for threshold selection and prove a general cross-validation result that justifies this approach.
\subsubsection{Soft Thresholding Estimator}
 \cite{antoniadis2001regularization} propose the Soft Thresholding method (ST) for covariance estimation takes off-diagonal elements $z$ of sample covariance matrix and applies
\[h_{\tau}(z) = sgn(z)(|z| - \tau )_{+}\]
where $sgn(z)$ is a sign of the value $z$, and $(x)_{+} = max(x, 0)$. If thr is rather a vector of regularization parameters, it applies cross-validation scheme to select an optimal value.
\subsubsection{Adaptive Thresholding}
 \cite{cai2011adaptive} proposed an adaptive variant (AT) of Bickel and Levina (2008). The
idea of adaptive thresholding is to apply thresholding technique on correlation matrix in that it becomes adaptive in terms of each variable. The adaptive estimator is $\hat{\Sigma}^{*} (\hat{\sigma}^{*}_{ij})_{p \times p}$
where 
\[\hat{\sigma}^{*}_{ij} = s_{\lambda_{ij}}(\hat{\sigma}_{ij})\]
where $s_{\lambda}(z)$ is a general thresholding function similar to those used in \cite{cai2011adaptive}.

\subsection{Random Matrix Theory Methods}
 RMT methods aim at distinguishing signal from noise, i.e. eigenvectors and eigenvalues of the correlation matrix containing real information (which one would like to include for risk control), from those which are devoid of any useful information and are unstable. From this point of view, it is interesting to compare the properties of an empirical correlation matrix
C to a \emph{null hypothesis} purely random matrix as one could obtain from a finite time series of strictly independent assets. Deviations from the random matrix case might then suggest the presence of true information. This approach is then used to clean and estimate large correlation matrices. We discuss one such approach suggested by \cite{bun2016rotational} below.  
\subsubsection{Rotationally Invariant Estimators}
\cite{bun2016rotational} propose an estimator $\Gamma(M)$  of the true signal $C$ which depends on the dataset $M$, where $M$ is the estimated correlation matrix. The true matrix $C$ is unknown and there are no particular insights on its components (the eigenvectors). Therefore they
construct their  estimator $\Gamma(M)$  in a rotationally invariant way from the noisy observation $M$ that they have. In simple terms, this means that
there is no privileged direction in the N-dimensional
space that would allow one to bias the eigenvectors of
the estimator $\Gamma(M)$  in some special directions. More
formally, the estimator construction must obey
\[\Omega \Gamma(M)\Omega' = \Gamma(\Omega M \Omega' )\]
for any rotation matrix $\Omega$. Any estimator satisfying
the above equation is referred as a Rotational Invariant
Estimator (RIE). The best estimator  is the solution of the following
minimization problem
\[\Gamma(M) =\arg \min_{RI \Gamma(M)} ||C - \Gamma(M)||\]
considered over the set of all possible RI estimators
$\Gamma(M)$. Based on this problem, they propose an Oracle estimator for large correlation matrices (\cite{bun2017cleaning}) which, in the large $N$ limit, and for a large class of problems, can be expressed in terms of the (observable) limiting spectral density (LSD) of $M$ only. This cleaned correlation matrix can then be used to estimate the covariance matrix and its inverse. \cite{bun2017cleaning} also show its application in portfolio allocation and Principal Components Analysis.

\textbf{}
\subsection{Hyperparameter Tuning for GGM Methods}
All the GGM methods we discuss in this paper estimate the precision matrix based on the choice of parameters. The choice of parameters determines the sparsity pattern and other properties of the estimated precision matrix. We employ the Grid search approach to tune the parameters involves defining a set of values for each parameter and testing all possible combinations. All the GGM methods except Greedy Prune are single parameter methods where the parameters can assume continuous values. Greedy Prune is a two-parameter method, where one parameter can assume continuous values and the other parameter can assume only discrete values. 
\subsubsection{Crossvalidation Method I (CV1)}\label{Crossvalidation Method - I}
From a grid of parameters, we choose (using 5-fold crossvalidation) that parameter which minimizes the portfolio variance. The chosen parameter estimates the precision matrix, which is then sued to compute the portfolio variance. 
\[\mathbb{E}(\sigma^{2}) = \frac{1}{m_{holdout}}\sum_{k=1}^{m_{holdout}} var(X. w_{min})\]
where $w_{min}$ denotes the minimum variance portfolio weights. 
In our study, both the cross-validations are performed on the matrix of returns, which then yields the parameter with the least cross-validation error.
\subsubsection{Crossvalidation Method II (CV2)} \label{Crossvalidation Method - II}
We selected parameters using a 5-fold crossvalidation with
the following least-squares style crossvalidation objective, after standardizing the coordinates to each having
empirical variance 1 and mean 0:
\[E(\hat{\Theta)} = \frac{1}{nm_{holdout}}\sum_{i=1}^{n}\sum_{k=1}^{m_{holdout}}(X_{i}^{(k)} + \sum_{j \neq i}\frac{\hat{\Theta}_{ij} + \hat{\Theta}_{ji}}{2\hat{\Theta_{ii}}}X_{i}^{(k)})^{2}\]
The true $\Theta$ minimizes this objective as $m_{holdout} \to \infty$, making it equal to the sum of conditional variances; when the initial variances are set to 1, this objective simply measures the average amount of variance reduction achieved over the coordinates. \par
We treat the cross-validation objective functions discussed in \ref{Crossvalidation Method - I} and \ref{Crossvalidation Method - II} as the functions to be optimized. Since we do not know the gradient behaviour of these functions, We use a grid ranging from $(0,\infty)$ as the range over which a Nelder-Mead Optimization exercise is performed. This extension of the grid-choice accounts for both local and global extrema, which makes it possible to remove the bias in model performance due to a limited grid size.
\subsection{Hyperparameter Tuning of Shrinkage and Thresholding Methods}
The estimation problem for shrinkage estimators involves the choice of shrinkage constant, which determines the weightage of the sample covariance matrix and the target matrix in the shrinkage estimate. 
\begin{itemize}
\item The Ledoit Wolf linear estimator (LWL) arrives at the optimal shrinkage constant (or the shrinkage parameter) by considered a general asymptotics framework where the dimension size $p$ is allowed to tend to infinity as the sample size tends to infinity. The optimal shrinkage constant then tends to a limiting constant. This estimated optimal shrinkage constant is chosen as the parameter for LWL.
\item The RBLW approach provides an analytical formula, by applying the Rao-Blackwell theorem to the LWL method. Using several nontrivial Haar integral computations, it  obtains a simple closed form solution. On the other hand, the OAS method provides an iterative method to calculate the optimal shrinkage intensities.Beginning with an initial naive choice, each iteration is defined as the oracle solution when the unknown covariance is replaced by its estimate obtained in the previous iteration. 
\item The generalized linear shrinkage estimator (BDL) uses results from RMT to derive asymptotic deterministic equivalents of the optimal shrinkage intensities.
\item The analytical nonlinear estimator  (LWNL)follows a similar asymptotic methodology like \cite{ledoit2004well} to estimate the optimal shrinkage parameter athough it adapts the selection criteria of the parameter to the portfolio variance and sharpe ratio estimators.
\end{itemize}
The Hard Thresholding Estimator uses a five-fold cross validation approach to select the threshold based on an objective function which aims to reduce the estimation error of the covariance matrix. The Soft and Adaptive Thresholding methods follow a similar approach of hyperparameter selection. We employ the above methods of hyperparameter selection for the corresponding shrinkage and thresholding estimators to estimate the covariance matrix.
\section{Statistical Testing Procedures for Predictive Ability}\label{SPA and MCS}
The availability of several alternative model specifications being able to adequately describe the unobserved data generating process (DGP) opens the question of selecting the \emph{best fitting model} according to a given optimality criterion. In our study we want to address two questions: First, We try to evaluate whether a particular method can be chosen as a benchmark method, by evaluating whether is significantly outperformed by other methods. Second, we try to find if it also possible to provide a relative ranking of these competing methods. We perform a series of rejections of the competing methods till we reach a small set of methods which cannot be pruned any further. We use two approaches suggested by \cite{hansen2005test} and \cite{hansen2011model} correspondingly to address these questions. \newline
\newline
\textbf{Test for Superior Predictive Ability (SPA):} \newline
 Let $k = 0, . . . , l$ be the models being considered, with $k = 0$ being the chosen benchmark model and $k = 1, \ldots, l$ are the models the benchmark is being compared against. Each model leads
to a sequence of daily losses,$ L_{,t}$
, $t = 1,\ldots,n$.
The relative performance variables are defined as
\newcommand{\vect}[1]{\boldsymbol{#1}}
\[\vect{X_{k,t}} = L_{0,t} - L_{k,t}, \\\ k = 1, \ldots, l \\\ t = 1, . . . , n\]
Let $\vect{X_{t}} = (X_{1,t},\ldots, X_{l,t})'$ be a vector of relative performances and if $\mu = E(X_{t})$, the null hypothesis is
\[H_{0} : \mu \leq \vect{0}\]
that is, the benchmark model is not inferior to any of the alternative models when the objective is to minimize the expectation of the loss function considered. For a number of competing models and a model chosen ex-ante as the benchmark model, it tells whether the chosen model has been rightly considered as a benchmark model. in this context, it is important to note that we use the Portfolio Variance as the loss function for comparing the benchmark model against other models. To estimate the distribution of the test statistic used for the SPA test, we use the stationary bootstrap procedure of \cite{politis1994stationary}. This also requires testing for the stationarity of the data. We perform the Augemented Dickey Fuller (ADF) and the Kwiatkowski-Phillips-Schmidt-Shin (KPSS) tests for detecting stationarity. We find that the data (both daily and intra-day) is stationary. This makes our data suitable for using SPA and the Model Confidence Set Procedure. In our study, we use this to arrive at an acceptable level of certainty about the performance rankings of Glasso and Greedy Prune. On the other hand, while this test can give insights about the choice of benchmark models, we are still interested in finding a reliable performance ranking when there are a large number of models present. \newline
\newline
\textbf{Model Confidence Set Procedure:}\newline
 Hansen’s Model Confidence Set (MCS) procedure consists of a sequence of statistic tests which permits to construct a set of  \emph{Superior Set
Models} (SSM), where the null hypothesis of Equal Predictive Ability (EPA) is not rejected at a certain confidence level. The EPA test statistic is calculated for an arbitrary loss function that satisfies general weak stationarity conditions. Formally, let $Y_{t}$ be the observation at time
$t$ and $\hat{Y_{i,t}}$
the output of model $i$ at time $t$, the loss function $l_{i,t}$,t associated to the $i$–th model is defined as:
\begin{equation}
    l_{i,t} = l(Y_{t},\hat{Y_{i,t}})
\end{equation}

$d$ measures the difference between the output $\hat{Y_{i,t}}$ and the \textit{a posteriori} realisation $Y_{t}$. In our study, the portfolio variance $\sigma^{2}_{i,t}$ is used as loss function. Thus $\sigma^{2}_{i,t}$ represents the model output $Y_{i,t}$ in this case. \cite{jain2019can} show that the portfolio variance  can also be used as a loss function. In our case, we only rely on the estimate $\hat{Y_{i,t}}$ (portfolio variance, in our case) for the loss function. This is because the estimated portfolio variance is directly related to the estimated error and variance itself being an estimated quantity, there cannot be any \textit{a postetiori} realization for it. \par
We now briefly describe how the MCS procedure is performed. It begins with  an initial set of models $\hat{M}^{0}$ of dimension $m$  accounting for all the model specifications described for a given confidence level $1 - \alpha$, a smaller set, the superior set of
models, SSM, $\hat{M}^{*}_{1 -\alpha}$ of dimension $m^{*} \leq m$. The best scenario is when the final set comprises
of a single model, i.e. $m^{*} = 1$. Formally, let $d_{ij,t}$ denotes the loss differential between models
$i$ and $j$:
\begin{equation}
    d_{ij,t} = l_{i,t} - l_{j,t},  \\\ i,j = 1,\ldots, m, \\\ t = 1,\ldots, n
\end{equation}

and let
\begin{equation}
  d_{i,t} = (m-1)^{-1}\sum_{j \in M}d_{ij,t}, \\\ i = 1,\ldots, m   
\end{equation}
be the simple loss of model $i$ relative to any other model $j$ at time $t$. The EPA hypothesis
for a given set of models M can be formulated in two alternative ways:
\begin{align}\label{eq8}
     H_{0,M} : c_{i,j} = 0,   \quad i, j = 1, 2, . . . , m \nonumber \\ 
    H_{A,M} : c_{i,j} \neq 0, \quad i, j = 1, 2, . . . , m 
\end{align}
or 
\begin{align}\label{eq9}
     H_{0,M} : c_{i.} = 0,   \quad i  = 1, 2, . . . , m \nonumber \\ 
    H_{A,M} : c_{i.} \neq 0, \quad i = 1, 2, . . . , m 
\end{align}

where $c_{ij} = \mathbb{E} (d_{ij})$ and $c_{i.} = \mathbb{E}(d_{i.})$ are assumed to be finite and not time dependent. According to Hansen et al. (2011), in order to test the two hypothesis above, the following two statistics are constructed:
\begin{equation}\label{eq10}
  t_{ij} = \frac{\Bar{d_{ij}}}{\sqrt{\hat{Var}(d_{ij})}} \\\ \textrm{and} t_{i.} = \frac{\Bar{d_{i,.}}}{\sqrt{\hat{Var}(d_{i,.})}} \\\ \textrm{for}\\\ i,j \in M 
\end{equation}

where $\Bar{d_{i,.}} = (m - 1)^{-1} \sum_{j \in M}\Bar{d_{ij}}$ is the simple loss of the $i$-th model relative to the averages
losses across models in the set $M$, and $\Bar{d_{ij}} = (m - 1)^{-1} \sum_{ t= 1}^{m}d_{ij,t}$ measures the relative sample
loss between the $i$–th and $j$–th models, while $\hat{Var}(\Bar{d_{i,.}})$
and $\hat{Var}(\Bar{d_{i,j}})$
are bootstrapped estimates of $var(\Bar{d_{i,.}})$
and $var(\Bar{d_{i,j}})$ respectively. To calculate the bootstrapped variances $\hat{Var}(\Bar{d_{i,.}})$, we perform a block–bootstrap procedure of 5000 resamples, where the block length $p$ is the max number of significant parameters obtained
by fitting an $AR(p)$ process on all the $d_{ij}$ terms. The first t–statistic $t_{ij}$ is used in the well known test for comparing two forecasts (e.g. \cite{diebold2002comparing} and \cite{west1996asymptotic}), while the second one is used in \cite{hansen2005test} and \cite{hansen2011model}. As discussed 
 The two EPA null hypothesis presented in equations \ref{eq8} and \ref{eq9} map naturally into the two test statistics:
\begin{equation}\label{eq11}
    T_{R,M} = \max_{i,j \in M}|t_{ij}| \\\ \textrm{and} \\\ T_{max,M} = \max_{i \in M}t_{i.}
\end{equation}
where $t_{ij}$ and $t_{i.}$ are defined in equation\ref{eq10}. The test statistics defined in equation \ref{eq11} can
be used in order to test the hypothesis discussed above.
 Sequential testing procedures eliminate the worst model at each step, until the hypothesis of equal predictive ability
(EPA) is accepted for all the models belonging to the SSM. The test statistics used as an elimination criteria for the worst model are defined below:
\[e_{max,M} = \max_{i \in M}\frac{\Bar{d_{i,.}}}{\hat{Var}(d_{i,.})}  \quad{\textrm{and}} \\\\\ e_{R,M} = \max_{i} \{\sup_{j \in M}\frac{\Bar{d_{ij}}}{\sqrt{\hat{Var}(d_{ij})}}\}\]
respectively.\par
Summarizing, the MCS procedure to obtain the SSM, consists of the following steps:
\begin{itemize}
    \item Set $M = M_{0}$
    \item Test for EPA–hypothesis: if EPA is accepted terminate the algorithm and set $M^{*}_{1 -\alpha} = M$,otherwise use the elimination rules defined in equations to determine the worst model.
    \item Remove the worst model, and go to step 2.
\end{itemize}
The choice of $\alpha$ determines the number of models chosen as \emph{superior}. If the data is informative enough, the test returns only a single model as a superior model. In our analyses, we see that among a large class of models that we deal with, some of the algorithms are categorized as superior and some as inferior and for the superior set of models, the MCS method gives a ranking based on superior predictive ability. Thus, models classified as inferior suggest that the estimation error encountered while using them leads to relatively inferior predictive ability. In other words, the test helps us to determine methods which show lesser error in covariance or precision matrix estimation, which further translates into lower portfolio variance. 
\section{Synthetic Experiment - Design and  Results} \label{Synthetic Experiment}
We analyse the performance of the algorithms in terms of structure recovery and sample complexity. That is, we try to estimate the number of samples required for correctly estimating the nodes of a GGM network. The idea of \emph{correct} estimation is given by the samples needed to approximately recover true edge structure after thresholding using
the test $\frac{|\hat{\Theta_{ij}}|}{\sqrt{\hat{\Theta_{ii}}\hat{\Theta_{jj}}}} > \kappa/2$ , where $\kappa$ is the $\kappa$ for the true precision matrix (see \cite{kelner2020learning} for a detailed discussion). Apart from portfolio variance (which doubles up as loss function as discussed in the following sections), this approach also indicates the estimation accuracy of the GGMs under consideration. The lesser the number samples required, the more accurate is the method. \newline 
We consider a simple attractive GGM (with paths and cliques) and show that Greedy Prune and HybridMB always perform well compared to previous
methods and sometimes outperform them considerably. 
We set $d$, the maximum clique size (the maximum number of mutually interconnected nodes that a graph can have), as 5, such that $n/2$ (where $n$ represents the number of nodes or dimensions) is a multiple of $d$. Let $B$ be a standard Brownian motion in 1 dimension, and let  be the values of the $B$ at equally spaced points in the interval
$[1/2, 3/2]$ be given by: 
\[X_{1}, . . . , X_{n/2}\] that is, 
\[X_{1} = B(1/2), X_{2} = B((1/2) + 1/(n - 1)), \ldots  \] Equivalently, let the covariance matrix of this
block be $Cov(X_{i}, X_{j}) = 1/2+min(i, j)/n$, or take the Laplacian of the path and add the appropriate constant to the top-left entry.
Let the variables $X_{n/2+1},\ldots,X_{n}$
be independent of the Brownian motion, and let their precision matrix
be block-diagonal with $d \times d$ blocks of the form $\Theta_{1}$ where $\Theta_{1}$ is a rescaling of $\Theta_{0}$ so that the coordinates
have unit variance, and $\Theta_{0} = I - (\rho/d)~\vec{11}^{T}$
 where $\rho \in (0, 1)$. $\rho$ represents the edge strength of the graph, Trials and hyperparameter selection was performed to minimize the number of incorrect edges (with error of at most 0.25 incorrect edges per node). We compare Greedy Prune and HybridMB with Glasso, MB and  Clime.
 \begin{figure}[!h] 
  \centering
  \begin{subfigure}[b]{0.5\linewidth}
    \includegraphics[width=\linewidth]{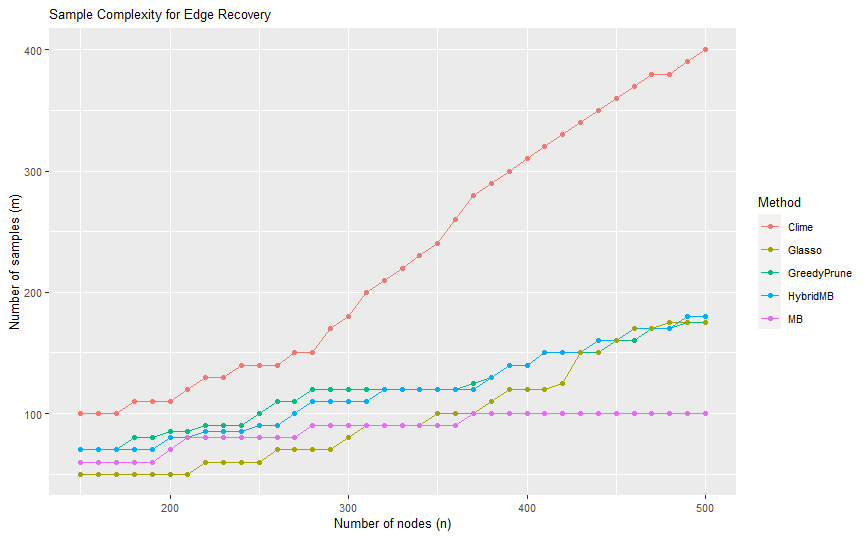}
    \caption{$d=5$, $\rho = 0.95$}
  \end{subfigure}
  \begin{subfigure}[b]{0.5\linewidth}
    \includegraphics[width=\linewidth]{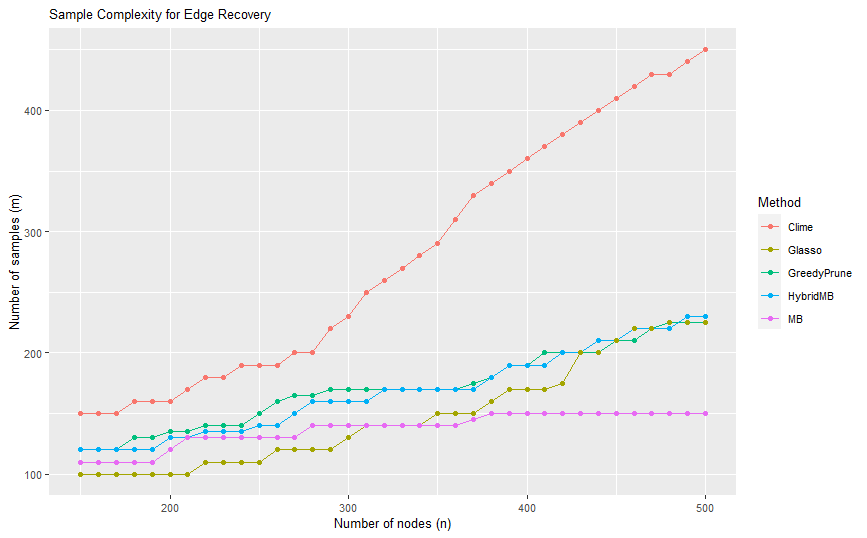}
    \caption{$d=5$, $\rho = 0.95 $}
  \end{subfigure}
  \caption{Synthetic Experiment: Number of samples needed to achieve error of at most 0.25 incorrect edges per node after thresholding: For Figures 1 a) and 1 b), CV1 and CV2 respectively were  used for tuning the parameters of the algorithms.}
 \label{Fig1}
\end{figure}
   Figure \ref{Fig1} depicts the behaviour of the algorithms when both the cross-validation schemes CV1 and CV2 are used for estimating the precision matrix respectively. $d= 5$ represents the maximum clique size of the graph. $\rho = 0.95$ denotes the edge strength of the graph. 
   
 We find that the Greedy Prune requires higher number of samples to estimate the nodes correctly compared to Glasso, but its sample requirements start catching up with Glasso as the number of nodes increases. What this means is that the sample complexity of Greedy Prune becomes better with the increase in the dimension size. Clime on the other hand, requires a much higher number of samples compared to all other algorithms. This holds true for the both the crossvalidation methods. The question that arises here is whether this also reflects in the precision matrix estimation error for Clime. This is answered in section \ref{Empirical Analysis - II}. HybridMB shows better sample complexity than Greedy Prune for lower number of nodes, but as the number of nodes increases, HybridMB shows a requirement of higher number of samples. This behaviour is observed till the number of nodes remains  
MB successfully achieved similar sample complexity to our proposed methods, despite the fact that this setup is again well outside of the regime where its theoretical guarantees are good. We see the sample complexity of Greedy Prune scales like O(log(n)). The results also suggest that the first cross-validation method (CV1) leads to lesser samples being required for node recovery compared to CV2.. Thus this experiment suggests the use of the minimum variance cross-validation scheme for node recovery or precision matrix estimation in the low sample regime. The difference in the predictive performance of CV1 and CV2 is further studied in the following sections.

We have already seen that Greedy Prune requires a similar number of samples compared to  Glasso in order to attain a similar level of structure recovery when the number of dimensions approaches 500. This might mean that Greedy Prune performs almost as well as Glasso in terms of accurate estimation of the precision matrix when the dimension size increases. To probe further, we performed an empirical experiment. We use the daily and intra-day returns of the constituent stocks of the Nifty500 index. We vary our sample sizes from 150 to 500 (with intervals of 50).The algorithms are applied using CV1. We computed the average out-of-sample portfolio variance (computed over rolling windows). 
\begin{figure}[!h] 
  \centering
    \includegraphics[width= 0.6\linewidth]{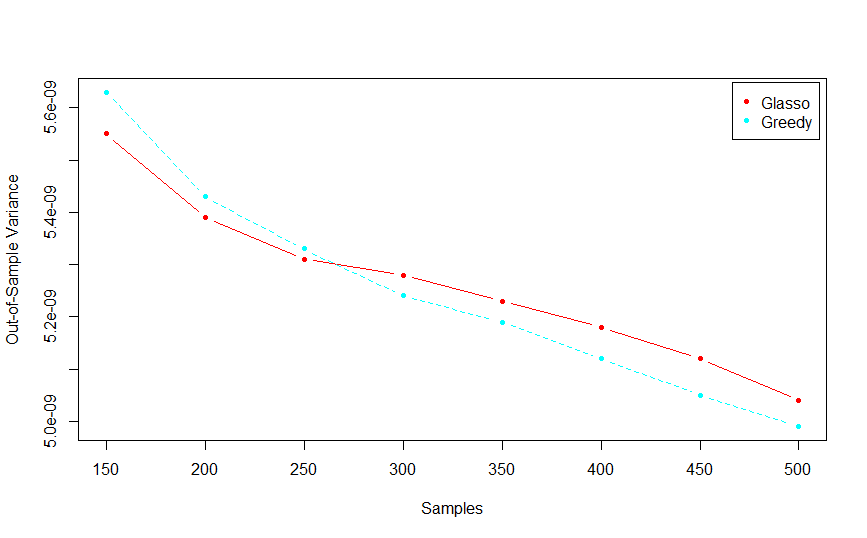}
    \caption{Out-of-Sample Variance Comparison for Glasso and Greedy Prune}
 \label{Fig2}
\end{figure}
In Figure \ref{Fig2} till 250 samples, Glasso outperforms Greedy.  After the number of samples approaches 250, Greedy starts outperforming Glasso. Thus, an increase in sample size, helps Greedy Prune attain an advantage (in terms of predictive ability) over Glasso. It is also important to note that a similar exercise performed for a dimension size of less than 50, Greedy Prune breaks down. This suggests that scaling up a high-dimensional problem improves the estimation ability of Greedy Prune. Also, when the dimension size is fixed, the increase in the number of samples also helps Greedy Prune to attain better estimation accuracy.

From the figure, it can be observed that Greedy Prune starts outperforming Glasso when the number of samples increase. The above experiment has been performed using the first cross-validation scheme (CV1).

 \section{Empirical Experiments: Data and Methodology}\label{Data and Methodology}
 To study how the algorithms discussed in section \ref{Data and Methodology} perform in terms of superior predictive ability in the context of the Global Minimum Variance portfolio, we use the Intra-Day and Daily Closing prices of the constituent stocks of the Nifty500 Index. We consider the time period from January 2015 to December 2022 (1724 days). The stock prices are use to compute the log-returns, which in turn are used in the portfolio optimization problem. In particular, we are interested in computing the portfolio variance using the inverse covariance matrices estimated using the techniques discussed above. This portfolio variance, as discussed by \cite{jain2019can}, can be used as an indicator of estimation error. That is, the higher the error in estimating the precision matrix, the more the portfolio variance. The properties (such as sparsity, condition number, etc.) of the precision matrix estimated using a given algorithm depends on the choice of the tuning parameter. The parameter is chosen using two five-fold cross validation methods, discussed in \ref{Crossvalidation Method - I} and \ref{Crossvalidation Method - II}.We consider three cases for portfolio rebalancing: 
(a) Daily $h = 1$, (b) weekly $h = 5$, and (c) monthly $h = 20$.
Further, we divide the daily estimation horizon into two periods: (a) Normal (2016-19) and (b) Stress (2020-2022). These two periods are considered to see whether the Covid-19 induced economic crisis impacts the correlation structure of the assets. This would then be possibly reflected in the relative performances of the algorithms. \par
We further perform an out-of-sample test to compare algorithm performance.
For the out-of-sample test procedure,  we use our set of algorithms to estimate the precision matrix for daily returns up until time $t$. We then use the intra-day returns matrix (of the $t+1$th day) to compute the portfolio variance.
\[ \underbrace{0,1,\ldots,\textrm{t}}_\textrm{training data} \\\ \underbrace{1,\ldots,\textrm{n}}_\textrm{t+1 - test data}\]
Thus, for a training window depicted above and data $X$, we estimate the covariance and precision matrices for all the competing methods, which finally yield us $w_{min,k,t}$ the minimum variance portfolio weights for a given method $k$ till period $t$. Using the intra-day returns ($X'$ for the next day we finally estimate the portfolio variance:
\[\sigma_{min, k, t+1} = X'.w_{min,k,t}\]
 For 500 stocks, we use a one-period overlapping rolling window of 150 days (this is the sample size), for daily horizon out-of-sample tests.
 
To obtain the weekly $t + 5$ and monthly $t + 20$ returns, we use the iterated sum of daily returns. 5 days of daily returns are summed up to get the weekly returns. 20 days of daily returns are summed up to get the monthly returns.For the weekly and monthly forecasting horizons, we use a rolling window of 100 weeks and 50 months respectively. This rolling window is use to calculate the optimal weights. We then use the next 5 and 20 days daily returns correspondingly to calculate the portfolio variances for the weekly and monthly horizons. Similarly, 20 days of daily returns are used to arrive at the monthly returns. Proceeding in this manner, we get a time series of losses for each method. This is then ready to be used for the SPA and MCS tests.
 
\section{Results for the Empirical Experiments} \label{Emp exp}
\subsection{Properties of Estimated Precision Matrices}\label{Empirical Analysis - I}
In this empirical experiment, we compare the summary statistics of the precision matrices ($\hat{\Theta}$ ) estimated using CV1 and CV2 for the Daily estimation Horizon (150 days, 500 stocks). The estimation process used here is In-Sample, that is the data used for tuning the hyperparameters is also used for calculating the portfolio variance.  These statistics give us an idea of the sparsity pattern, recovery error and other properties of the estimated precision matrix.  In Tables 1 and 2, the last 4 columns give summary statistics for the final recovered $\Theta$ using 5-fold CV on Nifty500 daily returns. The variables defined in the columns are as follows: \newline
\begin{itemize}
    \item Non-Zeros indicate the sparsity of the estimated precision matrix. Higher number of non-zero items indicate higher sparsity. 
    \item Cross-Validation Error indicates two things here. CV1 Error denotes the error incurred while selecting parameters according to the minimum variance of the portfolio. CV2 Error denotes the error incurred while selecting parameters by minimizing the variance per node.
    \item Condition Number is a measure of the sensitivity of a matrix to changes in its inputs. Here, a high condition number indicates that the matrix is ill-conditioned and that small changes in the input data can lead to large changes in the output. This can make the estimated precision matrix numerically unstable and result in inaccurate estimates. 
    \item The walk-summable relative error is $\Delta_{WS} := \frac{||\Tilde{\Theta} - \hat{\Theta}||_{F}}{||\hat{\Theta}||_{F}}$, where $\hat{\Theta}$ is the closest walk-summable matrix to $\Tilde{\Theta}$ in Frobenius norm. It indicates how close the estimated precision matrix to the theoretical walk-summable matrix. \footnote{More details on testing  for the walk-summability of an estimated precision matrix can be found in \cite{kelner2020learning}}.
\end{itemize}

\begin{table}[h!]
    \centering
    \begin{tabular}{|l|l|l|l|l|l|}
    \hline
        Method & CV1 Error  & Non-zeros & Condition No.   & Delta WS \\ \hline
        Glasso1 & 0.73  & 5.7E+4 & 41.62 &  6.15E-15 \\ \hline
        Clime1 & 0.87  & 2.3E+4 & 5.37 &  8.49E-15 \\ \hline
        Greedy Prune1 & 0.97  & 1.1E+4 & 16.02 &  1.38E-15 \\ \hline
        MB1 & 0.78 &  3.4E+4 & 8.52 &  6.18E-15 \\ \hline
        HybridMB1 & 0.94 &  3.1E+4 & 37.82 &  1.72E-14 \\ \hline
    \end{tabular}
    \caption{ Results for precision matrix selected via 5-fold CV1 on Nifty500 constituents}
    \label{t2}
\end{table}
In table \ref{t2}, we find that the all the estimated precision matrices are walk-summable. This essentially means the empirical joint distribution of the cross section of returns follows the assumption of walk-summability. This is interesting in the sense that it makes this data particularly suitable for applying GGM methods.  We also see from the first 2 columns
of the table, Graphical Lasso achieved the greatest amount of variance reduction but returned the densest estimate for $\Theta$, MB and HybridMB had slightly less portfolio variance reduction, Greedy Prune had the sparsest estimate but least variance reduction. We also see that the Greedy Prune, Clime, MB return the lowest condition numbers, thus indicating better stability of the estimates. Glasso1 and HybridMB have large condition numbers.

\begin{table}[h!] 
    \centering
    \begin{tabular}{|l|l|l|l|l|l|l|}
    \hline
        Method & CV2 Error & Non-zeros & Condition No. & Delta WS \\ \hline
        Glasso2 & 0.92 & 5.8E+4 & 47.56 &  7.15E-15 \\ \hline
        Clime2 & 1.01 & 2.5E+4 & 19.32 &  8.19E-15 \\ \hline
        Greedy Prune2 & 1.11  & 1.8E+4 & 34.42 &  3.16E-15 \\ \hline
        MB2 & 0.96 & 4.1E+4 & 21.69 & 6.72E-15 \\ \hline
        HybridMB2 & 1.08 & 3.1E+4 & 41.85 & 3.87E-14 \\ \hline
    \end{tabular}
    \caption{Results for precision matrix selected via 5-fold CV2 on the Nifty500 constituents}
    \label{t3}
\end{table}
We observe similar results for CV2 in table \ref{t3}. In comparison with  CV2, CV1 returns sparser $\Theta$ estimates. The condition numbers of the precision matrices estimated using CV2 is higher, which indicates greater sensitivity to noise. The last column again shows that that walk-summability is observed for all the methods under consideration. This further supports the observations in the previous table and provides a case for using GGM methods to study historical stock returns. A pattern clearly observed between CV1 and CV2 is the possibility of encountering higher noise in estimated precision matrices calculated using CV2. The question that follows is: Does it have an implication on portfolio allocation as well ? We look at this question in more details while studying GGM methods along with other models in the subsection that follows. 
\subsection{Model Comparison for the Minimum Variance Portfolio} \label{Empirical Analysis - II}
As mentioned previously, we perform an empirical experiment \footnote{Data and Code for this and some other experiments in this paper are available at: \href{https://github.com/sumanjaydutta/Gaussian-Graphical-Models_Codes}{Codes for Min-Variance Portfolio Study}} to compare the predictive performance of all the covariance and precision matrix estimation approaches discussed earlier in the context of Markowitz's portfolio allocation model. We perform this experiment to see whether estimation errors encountered while computing the inverse covariance matrix is related to the out-of-sample predictive ability of the algorithms under consideration. We divided our analyses into three estimation horizons: Daily, Weekly and Monthly. We compare the competing methods using MCS. Apart from this, we also consider the equally weighted portfolio (EWP) allocation scheme. The GGM algorithms are compared using using two cross-validation methods. Thus, we have 10 GGM methods \footnote{Algorithms tuned using CV1 are denote as Glasso1, Clime1, MB1, Greedy1 and HybridMB1. Algorithms tuned using CV2 are denote as Glasso2, Clime2, MB2, Greedy2 and HybridMB2} (5 models tuned in two different ways), 5 Shrinkage methods, 3 Thresholding methods (listed in \ref{Models}), the RIE estimator (based on RMT) and the EWP model for our comparison. Thus, we have 20 models in total, which makes the setup well suited for using the MCS procedure. \par
As mentioned previously, the Model Confidence Set procedure (MCS) begins with rejecting the inferior set of models and thereafter ranks the superior set of models. There are two ranking schemes as mentioned in Section \ref{SPA and MCS}. The models are then ranked according to their predictive ability. As mentioned earlier, MCS assumes EPA as a null hypothesis. The results for the three estimation horizons are presented below: 
\subsubsection{Comparison of Methods for the Daily Estimation Horizon}
As mentioned before, we consider a rolling window of 150 days (for 500 stocks) for out-of-sample tests. 
\begin{table}[htbp]
    \centering
    \begin{tabular}{|l|l|l|l|l|l|l|l|}
    \hline
       Model & $Rank_{M}$ & $v_{M}$ & $MCS_{M}$ & $Rank_{R}$ & $v_{R}$ & $MCS_{R}$\\ \hline
        Greedy1 & 1 & -1.51 & 1 & 3 & 0.19 & 1 \\ \hline
        HMB1 & 2 & -1.32 & 1 & 2 & 0.07 & 0.99 \\ \hline
        Glasso1 & 3 & -0.36 & 1 & 1 & -0.09 & 0.99 \\ \hline
        MB1 & 4 & 0.79 & 0.6 & 4 & 1.53 & 0.97  \\ \hline
        Clime1 & 5 & 1.98 & 0.5 & 5 & 3.12 & 0.91 \\ \hline
    \end{tabular}
     \caption{Out-of-Sample MCS test for Daily Estimation Horizon (Normal Period)}
      \label{t5}
\end{table}
We find that for the normal period (2015-19), the algorithms which use conditional variance minimizing cross-validation method are rejected and form the inferior set of models. Along with them, the covariance matrix estimation methods are rejected as well. This also includes the EWP. 
 Among the superior set of models, we find that Greedy Prune, HybridMB and Glasso perform the best. \newline
 The  subscripts $M$ and $R$ refer to the elimination rules associated with the test statistics defined in equation \ref{eq10}. The first test  statistic is associated with $\hat{d_{ij}}$,
measures the sample loss
differential between models $i$ and $j$. The second statistic  is associated with
$\hat{d_{i·}}$, which is a contrast of model $i$’s sample loss to that of
the average across all models. If for example, the second test statistic in \ref{eq10} is used, then the elimination rule is chosen such 
that it removes the model that contributes most to the test statistic among
the models with a sample performance that is worse than the average across models. Thus, Clime which is ranked 5th according to this rule,  
 has the largest standardized excess loss, relative to the average across
all models in M. In comparison Glasso, HybridMB and Greedy Prune have the lowest losses. The explanation for the rankings based on the second elimination rule is similar. 

To see how the covariance estimation methods perform among themselves, we perform a similar experiment
We present the performance ranking of Shrinkage, Thresholding and RMT Methods below:
\begin{table}[!h]
    \centering
    \begin{tabular}{|l|l|l|l|l|l|l|l|}
    \hline
       Model & $Rank_{M}$ & $v_{M}$ & $MCS_{M}$ & $Rank_{R}$ & $v_{R}$ & $MCS_{R}$ \\ \hline
        LWNL & 1 & -1.40 & 1 & 1 & -0.21 & 1 \\ \hline
        BDL  & 2 & -1.25 & 1 & 2 & -0.05 & 0.95 \\ \hline
        LWL   & 3 & 0.52 & 1 & 3 & 0.08 & 0.87 \\ \hline
        RBLW & 4 & 0.63 & 0.69 & 4 & 1.29 & 0.81 \\ \hline
        RIE  & 5 & 0.82 & 0.55 & 5 & 3.09 & 0.70 \\ \hline
    \end{tabular}
    \caption{Out-of-Sample MCS Test for Daily Estimation Horizon (Normal Period) - for Covariance Estimation Methods}
     \label{t6}
\end{table}
 In table \ref{t6}, we find that LWNL performs the best among all other models, followed by the linear shrinkage methods and the RIE. The thresholding methods are again classified as inferior models. This means that thresholding methods are clearly outperformed by shrinkage methods. We again compare all the methods for the stress period:
\begin{table}[htbp]
    \centering
    \begin{tabular}{|l|l|l|l|l|l|l|l|}
    \hline
       Model & $Rank_{M}$ & $v_{M}$ & $MCS_{M}$ & $Rank_{R}$ & $v_{R}$ & $MCS_{R}$  \\ \hline
        Greedy1 & 1 & -2.42 & 1 & 3 & -1.51 & 1 \\ \hline
        HMB1 & 2 & -1.83 & 1 & 2 & -0.45 & 0.99  \\ \hline
        Glasso1 & 3 & -1.25 & 1 & 1 & -0.41 & 0.95 \\ \hline
        MB1 & 4 & 0.65 & 0.49 & 4 & 1.22 & 0.85  \\ \hline
        Clime1 & 5 & 1.79 & 0.39 & 5 & 2.05 & 0.82  \\ \hline
    \end{tabular}
    \caption{Out-of-sample MCS test for Daily Estimation Horizon (Stress period)}
     \label{t7}
\end{table}
\newline
For the Stress period (table \ref{t7}), we find a similar ranking of the algorithms. Glasso, Greedy Prune and HybridMB are the best performers. Algorithms using the conditional variance minimizing cross-validation form the inferior set of models. While periods of economic and financial stress have been shown to indicate positive multicollinearity and high volatility, our stress-period (2020-22) does not indicate any impact on the relative performance of the algorithms. The covariance estimation methods are again classified as inferior models. Their rankings are the same as that for the normal period. This indicates that the stress period did not have an impact on the relative performances of the covariance matrix estimation methods as well. Shrinkage and RMT methods also continue to outperform thresholding approach. Among shrinkage methods, continuous to outperform linear shrinkage estimators. The next question that arises is related to the relative performance of all the methods for longer forecasting horizons, viz. weekly and monthly.

\subsubsection{Comparison of Methods for the Weekly Estimation Horizon}
 In this section, we present the results for the weekly portfolio estimation horizon. We consider the same portfolio set (500 stocks) and an estimation window of 100 weeks.  
    \begin{table}[h!]
    \centering
    \begin{tabular}{|l|l|l|l|l|l|l|l|}
    \hline
       Model & $Rank_{M}$ & $v_{M}$ & $MCS_{M}$ & $Rank_{R}$ & $v_{R}$ & $MCS_{R}$ \\ \hline
        Greedy1 & 1 & -2.19 & 1 & 3 & -0.52 & 1 \\ \hline
        HMB1 & 2 & -2.01 & 1 & 2 & -0.39 & 0.99 \\ \hline
        Glasso1 & 3 & -1.92 & 0.91 & 1 & -0.14 & 0.95  \\ \hline
         MB1 & 4 & -0.47 & 0.85 & 4 & 0.95 & 0.91 \\ \hline
        Clime1 & 5 & 0.54 & 0.83 & 5 & 2.42 & 0.75 \\ \hline
    \end{tabular}
    \caption{Out-of-Sample MCS Test for Weekly Estimation Horizon}
     \label{t8}
    \end{table}
Table  \ref{t8} presents the Out-of-Sample test for the weekly forecasting horizon. We find that the ranking of the algorithms remains the same. Also, the portfolio minimizing crossvalidation approach still outperforms the conditional variance minimizing crossvalidation scheme (which is classified as inferior). Similar to the previous case (daily), we find that EWP and the covariance estimation methods are categorized as inferior models. We perform the MCS test again for the covariance estimation methods, the results for which are given below 
\begin{table}[!h]
    \centering
    \begin{tabular}{|l|l|l|l|l|l|l|l|}
    \hline
       Model & $Rank_{M}$ & $v_{M}$ & $MCS_{M}$ & $Rank_{R}$ & $v_{R}$ & $MCS_{R}$  \\ \hline
         LWNL & 1 & -2.34 & 1 & 1 & -0.82 & 1 \\ \hline
        BDL & 2 & -1.87 & 1 & 2 & -0.35 & 0.99 \\ \hline
        LW  & 3 & -1.45 & 1 & 3 & -0.11 & 0.99 \\ \hline
        RIE & 4 & 0.42 & 0.97 & 4 & 0.05 & 0.95 \\ \hline
        RBLW  & 5 & 0.48 & 0.81 & 5 & 1.85 & 0.92 \\ \hline
    \end{tabular}
    \caption{Out-of-Sample MCS Test for Weekly Estimation Horizon - for  Covariance Matrix Estimation Methods }
     \label{t9}
\end{table}
In table \ref{t9}, we see that LWNL outperforms other methods, followed by BDL, LW, RIE and RBLW. Thresholding methods are further classified as relatively inferior models.
\subsubsection{Comparison of Methods for the Monthly Estimation Horizon}
 In this section, the results for the monthly estimation horizon.  
\begin{table}[h!]
    \centering
    \begin{tabular}{|l|l|l|l|l|l|l|l|}
    \hline
       Model & $Rank_{M}$ & $v_{M}$ & $MCS_{M}$ & $Rank_{R}$ & $v_{R}$ & $MCS_{R}$ \\ \hline
        Greedy1 & 1 & -1.64 & 1 & 3 & 0.02 & 0.97 \\ \hline
        HMB1 & 2 & -1.42 & 1 & 2 & -0.22 & 0.99  \\ \hline
        Glasso1 & 3 & -0.12 & 1 & 1 & -0.47 & 1 \\ \hline
        MB1 & 4 & 0.31 & 0.53 & 4 & 1.48 & 0.61  \\ \hline
        Clime1 & 5 & 0.70 & 0.39 & 5 & 3.12 & 0.12  \\ \hline
    \end{tabular}
     \caption{Out-of-Sample MCS Test for Monthly Estimation Horizon }
      \label{t10}
\end{table}
Table \ref{t10} presents Out-of-Sample test for the monthly forecasting horizon. We find that the ranking of the algorithms remains the same. Also, the GGM using portfolio minimizing cross-validation method still outperforms the conditional variance minimizing cross-validation scheme. Similar to the previous cases, we find that EWP, GGMs implemented using the CV2 scheme and  covariance matrix estimation methods are classified as inferior models. The rankings of the covariance matrix estimation methods are similar to the weekly estimation horizon with no particular change in the ordering of performance. Thresholding methods continue to be classified as inferior models.  \par
Our results for all the three forecasting horizons suggest that there is a conflict in the performance rankings of Glasso1 and Greedy Prune1 when different test statistics are used. This could possibly occur due to the choice of the test statistics. Hence we employ the SPA test of \cite{hansen2005test} to see whether the performance of Glasso1 and Greedy Prune1 remains similar and whether it is just the choice of the test statistic that contributes to the conflict in rankings.

\subsubsection{Assessing the performance of Glasso and Greedy Prune}
We choose the same set of methods as the one used in our empirical experiments for comparison. Glasso1 and Greedy Prune1 are chosen as the benchmark models for testing the null hypothesis. \ The results for the daily estimation horizon (normal period) are shown below:
\begin{table}[h!]
    \centering
    \begin{tabular}{|l|l|l|l|l|l|l|l|}
    \hline
       Model &  p-value \\ \hline
        Glasso1 &  0.953  \\ \hline
        Greedy1 & 0.974\\ \hline
    \end{tabular}
     \caption{Out-of-Sample SPA test for Daily Estimation Horizon - Normal period }
      \label{t11}
\end{table}
We find that both Glasso1 and Greedy1 can be chosen as benchmark models. Thus, both can be said to have comparable levels of predictive ability and this tells us that the anomaly in the rankings for the MCS test  was possibly only due to the choice of the test statistics. The observations are similar for weekly and monthly estimation horizons \footnote{Results are available with authors on request.}. 
\subsubsection{Estimation accuracy and Predictive Ability}
Our analysis involving the GGMs, EWP and the Covariance estimation methods in the previous sections shows that the GGMs implemented using the CV1 scheme outperform all other methods. Also, none of the Shrinkage, Thresholding and Random Matrix-based methods make it to the \emph{Superior} set of models. One possible hypothesis for this is that the precision matrices obtained by inverting the estimated covariance matrices from Shrinkage, Thresholding and Random Matrix-based  Methods are noisier when compare to the ones directly obtained from GGM methods. In order to probe into the possible reasons for such empirical observations, we perform a synthetic experiment. We generate synthetic data from a multivariate normal distribution where the underlying True Covariance matrix comes is calculated using the Nifty-500 dataset for daily returns using a sample size of 500. For 1000 repetitions, we estimate the precision matrices using four methods: Greedy Prune, Glasso, LWNL and BDL (the sample size is 150). The Frobenius ($L_{2}$) norm of the difference between the true precision matrix and the estimated precision matrices is computed. The results are presented below: 
\begin{table}[h!]
    \centering
    \begin{tabular}{|l|l|}
    \hline
       Model & Average Frobenius Error \\ \hline
        Greedy1 &  9.9E+4 (11.2) \\ \hline
        Glasso1 &  10.6E+4 (20.9) \\ \hline
        LWNL &  22.8E+4 (21.4)\\ \hline
        BDL & 23.E+4 (19.1)  \\ \hline
        
    \end{tabular}
     \caption{Frobenius Error for the best performing GGM and Shrinkage Methods}
     \label{t12}
\end{table}
The Parentheses in table \ref{t12} refer to the standard error of the $L_{2}$ distance between the true and the estimated precision matrices calculated for 1000 samples. Greedy1 yields lesser error  and is the most stable among all algorithms. It is followed by Glasso1, LWNL and BDL methods in terms of error reduction and stability.
\section{Conclusion}\label{Conclusion}
In this paper, we compare methods estimating the covariance and precision matrices on the basis of the estimation accuracy of the covariance and precision matrices and its implications on portfolio allocation. Our synthetic experiments show that the performance of the GGM algorithms is similar to that reported in \cite{kelner2020learning}. MB, Greedy Prune and HybridMB show better sample complexity among GGMs.  We also find that cross-validation scheme  proposed by us (CV1, see \ref{Crossvalidation Method - I}) for tuning GGM methods recovers the nodes  with fewer number of samples compared with \ref{Crossvalidation Method - II}.  This experiment on network learning also has implications for precision matrix. This is because the precision matrix reveals information about the network structure of a multivariate model. Hence, the extent of network learning should also be related to the estimation error for the precision matrix. The first empirical analysis studies the properties of the estimated precision matrices yielded by GGM methods. We see that CV1 yields sparser,ess erroneous and less sensitive estimates of the precision matrix compared to CV2. In the second empirical experiment, we compare all the methods in the context of Markowitz's minimum variance portfolio optimization model. We introduce portfolio variance as the loss function and employ the MCS test to compare different algorithms. We tune the GGM methods using both the schemes and find that CV1 enables superior predictive performance of the algorithms, compared to the CV2. GGM methods tuned using CV1 also outperform the covariance matrix estimation methods. Among the superior set of methods, Greedy, Prune, HybridMB and Glasso outperform MB and Clime.  Among the covariance matrix  estimation methods, we find LWNL to outperform the linear shrinkage estimators. Thresholding methods are clearly outperform by the shrinkage approach. RIE also performs better than the Thresholding methods. These observations generally hold true for all the forecasting horizons. We also saw that the Covid-19 pandemic did not impact the relaive performance of the algorithms. 

We find that for the two different test statistics, Glasso1 and Greedy Prune1 perform similarly well when the MCS test is used. we use the SPA test to find out whether both of them can be used as benchmark models. It is observed that both of them can be used as benchmark models and the change in the rank ordering of their relative performance (using MCS test) can be attributed to the test statistics under consideration. 
The comparison between GGMs and the best performing covariance estimation methods using a synthetic experiment also shows that estimation error for the precision matrix is indeed related to the performance rankings observed for the portfolio allocation problem. The paper also proposes a novel method of hyperparameter selection, which enables comparison of GGM methods using continuous grids.

\bibliographystyle{plainnat}
\bibliography{main}

\begin{thebibliography}{39}
\providecommand{\natexlab}[1]{#1}
\providecommand{\url}[1]{\texttt{#1}}
\expandafter\ifx\csname urlstyle\endcsname\relax
  \providecommand{\doi}[1]{doi: #1}\else
  \providecommand{\doi}{doi: \begingroup \urlstyle{rm}\Url}\fi

\bibitem[Antoniadis and Fan(2001)]{antoniadis2001regularization}
Anestis Antoniadis and Jianqing Fan.
\newblock Regularization of wavelet approximations.
\newblock \emph{Journal of the American Statistical Association}, 96\penalty0
  (455):\penalty0 939--967, 2001.

\bibitem[Awoye(2016)]{awoye2016markowitz}
Oluwatoyin~Abimbola Awoye.
\newblock \emph{Markowitz minimum variance portfolio optimization using new
  machine learning methods}.
\newblock PhD thesis, (UCL) University College London, 2016.

\bibitem[Bickel and Levina(2008)]{bickel2008covariance}
Peter~J Bickel and Elizaveta Levina.
\newblock Covariance regularization by thresholding.
\newblock \emph{The Annals of statistics}, 36\penalty0 (6):\penalty0
  2577--2604, 2008.

\bibitem[Bodnar et~al.(2014)Bodnar, Gupta, and Parolya]{bodnar2014strong}
Taras Bodnar, Arjun~K Gupta, and Nestor Parolya.
\newblock On the strong convergence of the optimal linear shrinkage estimator
  for large dimensional covariance matrix.
\newblock \emph{Journal of Multivariate Analysis}, 132:\penalty0 215--228,
  2014.

\bibitem[Bun and Knowles(2018)]{bun2018optimal}
Jo{\"e}l Bun and A~Knowles.
\newblock An optimal rotational invariant estimator for general covariance
  matrices: The outliers.
\newblock \emph{Preprint}, 2018.

\bibitem[Bun et~al.(2016)Bun, Allez, Bouchaud, and Potters]{bun2016rotational}
Jo{\"e}l Bun, Romain Allez, Jean-Philippe Bouchaud, and Marc Potters.
\newblock Rotational invariant estimator for general noisy matrices.
\newblock \emph{IEEE Transactions on Information Theory}, 62\penalty0
  (12):\penalty0 7475--7490, 2016.

\bibitem[Bun et~al.(2017)Bun, Bouchaud, and Potters]{bun2017cleaning}
Jo{\"e}l Bun, Jean-Philippe Bouchaud, and Marc Potters.
\newblock Cleaning large correlation matrices: tools from random matrix theory.
\newblock \emph{Physics Reports}, 666:\penalty0 1--109, 2017.

\bibitem[Cai et~al.(2020)Cai, Hu, Li, and Zheng]{cai2020high}
T~Tony Cai, Jianchang Hu, Yingying Li, and Xinghua Zheng.
\newblock High-dimensional minimum variance portfolio estimation based on
  high-frequency data.
\newblock \emph{Journal of Econometrics}, 214\penalty0 (2):\penalty0 482--494,
  2020.

\bibitem[Cai and Liu(2011)]{cai2011adaptive}
Tony Cai and Weidong Liu.
\newblock Adaptive thresholding for sparse covariance matrix estimation.
\newblock \emph{Journal of the American Statistical Association}, 106\penalty0
  (494):\penalty0 672--684, 2011.

\bibitem[Cai et~al.(2011)Cai, Liu, and Luo]{cai2011constrained}
Tony Cai, Weidong Liu, and Xi~Luo.
\newblock A constrained $\ell_{1}$ minimization approach to sparse precision
  matrix estimation.
\newblock \emph{Journal of the American Statistical Association}, 106\penalty0
  (494):\penalty0 594--607, 2011.

\bibitem[Callot et~al.(2021)Callot, Caner, {\"O}nder, and
  Ula{\c{s}}an]{callot2021nodewise}
Laurent Callot, Mehmet Caner, A~{\"O}zlem {\"O}nder, and Esra Ula{\c{s}}an.
\newblock A nodewise regression approach to estimating large portfolios.
\newblock \emph{Journal of Business \& Economic Statistics}, 39\penalty0
  (2):\penalty0 520--531, 2021.

\bibitem[Chen et~al.(2010)Chen, Wiesel, Eldar, and Hero]{chen2010shrinkage}
Yilun Chen, Ami Wiesel, Yonina~C Eldar, and Alfred~O Hero.
\newblock Shrinkage algorithms for mmse covariance estimation.
\newblock \emph{IEEE Transactions on Signal Processing}, 58\penalty0
  (10):\penalty0 5016--5029, 2010.

\bibitem[Diebold and Mariano(2002)]{diebold2002comparing}
Francis~X Diebold and Robert~S Mariano.
\newblock Comparing predictive accuracy.
\newblock \emph{Journal of Business \& economic statistics}, 20\penalty0
  (1):\penalty0 134--144, 2002.

\bibitem[Donoho et~al.(1995)Donoho, Johnstone, Kerkyacharian, and
  Picard]{donoho1995wavelet}
David~L Donoho, Iain~M Johnstone, G{\'e}rard Kerkyacharian, and Dominique
  Picard.
\newblock Wavelet shrinkage: asymptopia?
\newblock \emph{Journal of the Royal Statistical Society: Series B
  (Methodological)}, 57\penalty0 (2):\penalty0 301--337, 1995.

\bibitem[Fan et~al.(2013)Fan, Liao, and Mincheva]{fan2013large}
Jianqing Fan, Yuan Liao, and Martina Mincheva.
\newblock Large covariance estimation by thresholding principal orthogonal
  complements.
\newblock \emph{Journal of the Royal Statistical Society: Series B (Statistical
  Methodology)}, 75\penalty0 (4):\penalty0 603--680, 2013.

\bibitem[Friedman et~al.(2008)Friedman, Hastie, and
  Tibshirani]{friedman2008sparse}
Jerome Friedman, Trevor Hastie, and Robert Tibshirani.
\newblock Sparse inverse covariance estimation with the graphical lasso.
\newblock \emph{Biostatistics}, 9\penalty0 (3):\penalty0 432--441, 2008.

\bibitem[Frost and Savarino(1986)]{frost1986empirical}
Peter~A Frost and James~E Savarino.
\newblock An empirical bayes approach to efficient portfolio selection.
\newblock \emph{Journal of Financial and Quantitative Analysis}, 21\penalty0
  (3):\penalty0 293--305, 1986.

\bibitem[Hansen et~al.(2011)Hansen, Lunde, and Nason]{hansen2011model}
Peter~R Hansen, Asger Lunde, and James~M Nason.
\newblock The model confidence set.
\newblock \emph{Econometrica}, 79\penalty0 (2):\penalty0 453--497, 2011.

\bibitem[Hansen(2005)]{hansen2005test}
Peter~Reinhard Hansen.
\newblock A test for superior predictive ability.
\newblock \emph{Journal of Business \& Economic Statistics}, 23\penalty0
  (4):\penalty0 365--380, 2005.

\bibitem[Jagannathan and Ma(2003)]{jagannathan2003risk}
Ravi Jagannathan and Tongshu Ma.
\newblock Risk reduction in large portfolios: Why imposing the wrong
  constraints helps.
\newblock \emph{The Journal of Finance}, 58\penalty0 (4):\penalty0 1651--1683,
  2003.

\bibitem[Jain and Jain(2019)]{jain2019can}
Prayut Jain and Shashi Jain.
\newblock Can machine learning-based portfolios outperform traditional
  risk-based portfolios? the need to account for covariance misspecification.
\newblock \emph{Risks}, 7\penalty0 (3):\penalty0 74, 2019.

\bibitem[Jorion(1986)]{jorion1986bayes}
Philippe Jorion.
\newblock Bayes-stein estimation for portfolio analysis.
\newblock \emph{Journal of Financial and Quantitative analysis}, 21\penalty0
  (3):\penalty0 279--292, 1986.

\bibitem[Kelner et~al.(2020)Kelner, Koehler, Meka, and
  Moitra]{kelner2020learning}
Jonathan Kelner, Frederic Koehler, Raghu Meka, and Ankur Moitra.
\newblock Learning some popular gaussian graphical models without condition
  number bounds.
\newblock \emph{Advances in Neural Information Processing Systems},
  33:\penalty0 10986--10998, 2020.

\bibitem[Ledoit and P{\'e}ch{\'e}(2011)]{ledoit2011eigenvectors}
Olivier Ledoit and Sandrine P{\'e}ch{\'e}.
\newblock Eigenvectors of some large sample covariance matrix ensembles.
\newblock \emph{Probability Theory and Related Fields}, 151\penalty0
  (1-2):\penalty0 233--264, 2011.

\bibitem[Ledoit and Wolf(2004{\natexlab{a}})]{ledoit2004honey}
Olivier Ledoit and Michael Wolf.
\newblock Honey, i shrunk the sample covariance matrix.
\newblock \emph{The Journal of Portfolio Management}, 30\penalty0 (4):\penalty0
  110--119, 2004{\natexlab{a}}.

\bibitem[Ledoit and Wolf(2004{\natexlab{b}})]{ledoit2004well}
Olivier Ledoit and Michael Wolf.
\newblock A well-conditioned estimator for large-dimensional covariance
  matrices.
\newblock \emph{Journal of multivariate analysis}, 88\penalty0 (2):\penalty0
  365--411, 2004{\natexlab{b}}.

\bibitem[Ledoit and Wolf(2020)]{ledoit2020analytical}
Olivier Ledoit and Michael Wolf.
\newblock Analytical nonlinear shrinkage of large-dimensional covariance
  matrices.
\newblock \emph{The Annals of Statistics}, 48\penalty0 (5):\penalty0
  3043--3065, 2020.

\bibitem[Malioutov et~al.(2006)Malioutov, Johnson, and
  Willsky]{malioutov2006walk}
Dmitry~M Malioutov, Jason~K Johnson, and Alan~S Willsky.
\newblock Walk-sums and belief propagation in gaussian graphical models.
\newblock \emph{The Journal of Machine Learning Research}, 7:\penalty0
  2031--2064, 2006.

\bibitem[Markowitz(1952)]{markowitz77portfolio}
Harry Markowitz.
\newblock Portfolio selection.
\newblock \emph{The Journal of Finance}, 7\penalty0 (1):\penalty0 77--91, 1952.
\newblock ISSN 00221082, 15406261.
\newblock URL \url{http://www.jstor.org/stable/2975974}.

\bibitem[Meinshausen and B{\"u}hlmann(2006)]{meinshausen2006high}
Nicolai Meinshausen and Peter B{\"u}hlmann.
\newblock High-dimensional graphs and variable selection with the lasso.
\newblock \emph{The annals of statistics}, 34\penalty0 (3):\penalty0
  1436--1462, 2006.

\bibitem[Meucci(2005)]{meucci2005risk}
Attilio Meucci.
\newblock \emph{Risk and asset allocation}, volume~1.
\newblock Springer, 2005.

\bibitem[Millington and Niranjan(2017)]{millington2017robust}
Tristan Millington and Mahesan Niranjan.
\newblock Robust portfolio risk minimization using the graphical lasso.
\newblock In \emph{International Conference on Neural Information Processing},
  pages 863--872. Springer, 2017.

\bibitem[Misra et~al.(2020)Misra, Vuffray, and Lokhov]{misra2020information}
Sidhant Misra, Marc Vuffray, and Andrey~Y Lokhov.
\newblock Information theoretic optimal learning of gaussian graphical models.
\newblock In \emph{Conference on Learning Theory}, pages 2888--2909. PMLR,
  2020.

\bibitem[Politis and Romano(1994)]{politis1994stationary}
Dimitris~N Politis and Joseph~P Romano.
\newblock The stationary bootstrap.
\newblock \emph{Journal of the American Statistical association}, 89\penalty0
  (428):\penalty0 1303--1313, 1994.

\bibitem[Romano and Wolf(2005)]{romano2005stepwise}
Joseph~P Romano and Michael Wolf.
\newblock Stepwise multiple testing as formalized data snooping.
\newblock \emph{Econometrica}, 73\penalty0 (4):\penalty0 1237--1282, 2005.

\bibitem[Sch{\"a}fer and Strimmer(2005)]{schafer2005shrinkage}
Juliane Sch{\"a}fer and Korbinian Strimmer.
\newblock A shrinkage approach to large-scale covariance matrix estimation and
  implications for functional genomics.
\newblock \emph{Statistical applications in genetics and molecular biology},
  4\penalty0 (1), 2005.

\bibitem[Torri et~al.(2019)Torri, Giacometti, and Paterlini]{torri2019sparse}
Gabriele Torri, Rosella Giacometti, and Sandra Paterlini.
\newblock Sparse precision matrices for minimum variance portfolios.
\newblock \emph{Computational Management Science}, 16\penalty0 (3):\penalty0
  375--400, 2019.

\bibitem[West(1996)]{west1996asymptotic}
Kenneth~D West.
\newblock Asymptotic inference about predictive ability.
\newblock \emph{Econometrica: Journal of the Econometric Society}, pages
  1067--1084, 1996.

\bibitem[White(2000)]{white2000reality}
Halbert White.
\newblock A reality check for data snooping.
\newblock \emph{Econometrica}, 68\penalty0 (5):\penalty0 1097--1126, 2000.

\end{thebibliography}

\appendix 
\section{Appendix 1: Analyzing the Precision Matrix using Spectral Decomposition}\label{Ap1}

 In the context of the input estimates for the Markowitz portfolio allocation, an important critique is related to the usage of the sample covariance estimates. The second shortcoming is related to the role of the inverse covariance matrix $\Theta \equiv \Sigma^{-1}$, in the solution of the optimization problem. The accuracy of the precision matrix estimate directly affects the accuracy of the estimation of the weights. The relationship between the two matrices can be explicitly studied using spectral decomposition. 
The eigenvector decomposition of the covariance matrix can be expressed
as $\Sigma = V\Lambda V,$
where $V$ is the matrix of eigenvectors with $V^{-1} = V'$
and $\Lambda = diag(\lambda_{1}, \ldots, \lambda_{n})$ with $\lambda_{1} \geq  \cdots \geq \lambda_{n}$ is the diagonal matrix of the eigenvalues sorted in decreasing order on the main diagonal. Analogously, the
eigendecomposition of the precision matrix is such that $\Theta = U\Delta U'$, where $\Delta = diag(\delta_{1},\ldots,\delta_{n})$ with $\delta_{1} \leq \cdots \leq \delta_{n}$. By inverting the covariance matrix, we have

  \[  \Sigma^{-1} = V\Lambda^{-1}V' \]

It follows that $U = V$ and $\Delta = \Lambda^{-1}$ with the ith element $\delta_{i} = \frac{1}{\lambda_{i}}$,that is, the eigenvectors of the precision matrix are the same as those of the covariance matrix and the eigenvalues are the reciprocal of those of the covariance matrix. This shows how errors encountered in estimating the covariance matrix can also impact the inverse of the estimated covariance matrix. 

\section{Appendix 2: Extending Grid Size using Nelder Mead Optimization}\label{Ap2}
The Graphical Lasso and other techniques are based on regularization techniques such as Lasso which make parameter choice from a finite sized grid. The Greedy Prune and HybridMB algorithms use two (one discrete and one continuous) and one parameters respectively. The literature on GGMs does not discuss the impact of parameter choice on the behaviour of the extrema (both local and global) of the objective functions. This makes comparison of algorithms heavily dependent on the choice of the range and size of the grids. Thus, there is a scope for addressing the possibility of the usage of a finite sized grids. We employ the Nelder Mead optimization method to address this issue. Current literature does not discuss the behaviour of the objective function over a range of grid values. Thus nothing can be inferred theoretically or empirically about the possibility of convexity, continuity, etc. 
The method does not require any derivative information, which makes it suitable for problems with non-smooth functions. It is widely used to solve parameter estimation and similar statistical problems, where the function values are uncertain or subject to noise. It can also be used for problems with discontinuous functions, which occur frequently in statistics and experimental mathematics.
\section{Appendix 3: List of Models}\label{Ap3}
Table \ref{t13} presents a list of the models studied in this paper.
\begin{table}[]
\begin{tabular}{|p{0.3\linewidth} | p{0.3\linewidth} | p{0.3\linewidth}|}
\hline
Model                     & Class                                         & Reference \\ \hline
 Glasso                    & GGM                &        \cite{friedman2008sparse}   \\ \hline
MB                        & GGM               & \cite{meinshausen2006high}          \\ \hline
 Clime                     & GGM                 &        \cite{cai2011constrained}   \\ \hline
 Greedy Prune               & GGM           & \cite{kelner2020learning}          \\ \hline
 HybridMB                  & GGM           & \cite{kelner2020learning}          \\ \hline
Ledoit Wolf Linear        & Shrinkage                                     &  \cite{ledoit2004well}       \\ \hline 
 Rao Blackwell Ledoit Wolf & Shrinkage                                     &       \cite{chen2010shrinkage}    \\ \hline
Oracle Approaching        & Shrinkage                                  &     \cite{chen2010shrinkage}      \\ \hline
  Bodnar Linear             & Shrinkage                                     &        \cite{bodnar2014strong}   \\ \hline
 Ledoit Wolf Non Linear    & Shrinkage                                     &      \cite{ledoit2020analytical}      \\ \hline
  Soft                      & Thresholding                                  &      \cite{antoniadis2001regularization}     \\ \hline
  Hard                      & Thresholding                                  &       \cite{chen2010shrinkage}    \\ \hline
 
  Adaptive                  & Thresholding                                  &      \cite{cai2011adaptive}     \\ \hline
 Rotationally Invariant Estimator            & Random Matrix Theory                          &       \cite{bun2016rotational}    \\ \hline
\end{tabular}
\caption{List of Estimation Techniques}
\label{t13}
\end{table}
\newpage

\end{document}